\documentclass[a4paper,fleqn]{cas-sc}
\usepackage{algorithm}
\usepackage{algpseudocode}
\usepackage{tabularx}
\newcolumntype{Y}{>{\centering\arraybackslash}X}
\usepackage{array}
\usepackage[numbers,sort&compress]{natbib}

\def\tsc#1{\csdef{#1}{\textsc{\lowercase{#1}}\xspace}}
\tsc{WGM}
\tsc{QE}
\tsc{EP}
\tsc{PMS}
\tsc{BEC}
\tsc{DE}

\begin{document}
\let\WriteBookmarks\relax
\def\floatpagepagefraction{1}
\def\textpagefraction{.001}

\shorttitle{Phase-field modeling of T1 precipitates}

\shortauthors{A. Safi et~al.}

\title [mode = title]{A multi-component phase-field model for T1 precipitates in Al-Cu-Li alloys}

%
\author[1]{{Ali Reza} Safi}[type=editor,
orcid=0000-0003-4467-6734]

\cormark[1]

\ead{ali.safi@hereon.de}

\credit{Conceptualization, Data curation, Formal analysis, Investigation, Methodology, Software, Visualization, Writing - original draft}

\affiliation[1]{organization={Helmholtz-Zentrum Hereon, Institute of Material and Process Design, Solid State Materials Processing},
     addressline={Max–Planck-Straße 1}, 
    citysep={}, 
    city={21502 Geesthacht},
    country={Germany}}

\affiliation[2]{organization={Leuphana University Lüneburg, Institute for Production Technology and Systems},
    addressline={Universitäatsallee 1}, 
    city={21335 Lüneburg},
    country={Germany}}
    
\author[1]{Elizabeth Mathew}[style=chinese]
\credit{Data curation, Methodology, Software, Writing - review \& editing}

\author[1]{Rupesh Chafle}[style=chinese]
\credit{Formal analysis, Project administration, Supervision, Writing - review \& editing}

\author[1,2]{Benjamin Klusemann}[style=chinese]
\credit{Funding acquisition, Project administration, Supervision, Writing - review \& editing}

\cortext[cor1]{Corresponding author}


\begin{abstract}
In this study, the role of elastic and interfacial energies in the shape evolution of T\textsubscript{1} precipitates in Al-Cu-Li alloys is investigated using phase-field modeling. We employ a formulation considering the stoichiometric nature of the precipitate phase explicitly, including coupled equation systems for various order parameters. Inputs such as elastic properties are derived from DFT calculations, while chemical potentials are obtained from CALPHAD databases. This methodology provides a framework that is consistent with the derived chemical potentials to study the interplay of thermodynamic, kinetic, and elastic effects on T\textsubscript{1} precipitate evolution in Al-Cu-Li alloys. It is shown that diffusion-controlled lengthening and interface-controlled thickening are important mechanisms to describe the growth of T\textsubscript{1} precipitates. Furthermore, the study illustrates that the precipitate shape is significantly influenced by the anisotropy in interfacial energy and linear reaction rate, however, elastic effects only play a minor role.
\end{abstract}


\begin{highlights}
\item A stoichiometric multi-component phase-field model is presented to predict the precipitate evolution of the T\textsubscript{1} phase in Al-Cu-Li.
\item CALPHAD methods are used to derive the chemical potentials of the phases while first-principles calculations are performed to predict bulk elastic properties of $\alpha$ and T\textsubscript{1}.
\item Diffusion-controlled lengthening and interface-controlled thickening significantly contribute to the anisotropic plate shape of T\textsubscript{1} precipitates.
\item Multi-particle simulations are performed to demonstrate the complex interaction mechanisms during the growth of different crystallographic variants of T\textsubscript{1}.
\end{highlights}


\maketitle

\section{Introduction}
Aluminum (Al) alloys derive their properties from a complex interplay of composition, microstructure, and thermo-mechanical processing. In particular, the formation of various precipitates during heat treatment processes is responsible for the unique characteristics of these alloys, such as strength, ductility, and corrosion resistance \cite{williamsProgressStructuralMaterials2003}. Precipitation hardening, a crucial phenomenon in Al alloys, involves the controlled formation and distribution of precipitates from a supersaturated solid solution. This process is influenced by factors such as alloy composition, aging temperature and time.

T$_1$ (Al$_2$CuLi) and $\theta'$ (Al$_2$Cu) precipitates are considered the most important strengthening contributors in Al-Cu and Al-Cu-Li alloys, enhancing the alloy's strength through mechanisms like Orowan looping \cite{orowanZurKristallplastizitaetIII1934, nieEffectPrecipitateShape1996} and particle shearing \cite{deschampsInfluencePredeformationAgEing1998}. The theory of precipitate strengthening contribution distinguishes between these mechanisms, while emphasizing the effects of precipitate size, distribution, and the matrix-precipitate lattice mismatch. In artificially aged Al alloys, dislocations accumulate more rapidly compared to identical solid solutions, forming geometrically-necessary dislocations that develop in arrays around the precipitates during plastic deformation \cite{russellSlipAluminumCrystals1970}. Plastic deformation prior to artificial aging can further promote the nucleation of T$_1$ precipitates, changing the relative volume fraction of T$_1$ to $\theta'$ \cite{gableRolePlasticDeformation2001}.

Most of the strongest conventional precipitation-hardened Al alloys feature thin plate-shaped precipitates, like T$_1$ and $\theta'$, that align with the primary slip planes of the $\alpha$-Al solid solution matrix \cite{nieEffectPrecipitateShape1996}. The strengthening contribution from these precipitates can be accurately modeled by employing analytical equations for precipitation strengthening or dispersion hardening to consider the precipitates' shape, orientation, and distribution \cite{nieMicrostructuralDesignHighstrength1998}. Recent shear strengthening models further distinguish between the hardening effects resulting from the intrinsic stacking fault energy in these precipitates and interfacial energy between precipitate and matrix \cite{dorinQuantificationModellingMicrostructure2014}.

Decreus et al. \cite{decreusInfluenceCuLi2013} described the structure of T$_1$ phase precipitates in detail, noting their layered structure and the minimal lattice mismatch with the $(111)$ Al planes, which allows these precipitates to grow to large dimensions without losing coherency. Recent studies of Häusler et al. \cite{hauslerPrecipitationT1Phase2017} have explored the age-hardening response of high-purity Al–4Cu–1Li–0.25Mn alloys. T$_1$ precipitates undergo a unique thickening process that is clearly distinguished from their lengthening behavior. This involves alternating stacking sequences of elementary structures identified as Type 1 and Defect Type 1, leading to the thickening of the T$_1$ precipitates. Type 1 structures are characterized by a Li-rich central layer, while Defect Type 1 structures lack this Li-rich layer. This mechanism differs from the growth of $\theta'$ precipitates, which follows a ledge growth mechanism. Further experimental studies have shown that holding an aging temperature of 155~$^{\circ}\mathrm{C}$ results in T$_1$ precipitates maintaining minimal thickness over long aging times. Increasing the temperature to 190~$^{\circ}\mathrm{C}$ rapidly activates the thickening of T$_1$ precipitates \cite{dorinQuantificationModellingMicrostructure2014}. This temperature-dependent behavior suggests that interface mobility of the $(111)_{\alpha} // (0001)_{\text{T}_1}$ interface, as well as solubility and diffusion rates of the alloying elements, are key factors that influence the thickening process.

While mean field modeling strategies, such as Kampmann-Wagner numerical methods \cite{herrnringDiffusiondrivenMicrostructureEvolution2020, herrnringModelingPrecipitationKinetics2021, uryKawinOpenSource2023}, can provide statistically relevant insights on particle size distribution and number density on a macroscopic scale, they lack the ability of providing a detailed understanding of precipitate shapes on smaller lengthscales. In contrast, mesoscale models, such as the phase-field method, have been demonstrated to be powerful tools to study precipitate evolution. It allows for the quantification of competing effects, such as strain energy, interfacial energy, and other thermodynamic driving forces, capturing the dynamics of precipitate growth \cite{chenPhaseFieldModelsMicrostructure2002, steinbachPhaseFieldModelMicrostructure2013, tourretPhasefieldModelingMicrostructure2022}. While there are extensive studies on the application of phase-field modeling on $\theta'$ \cite{jiPhasefieldModelStoichiometric2022, liuMultiscaleModellingMorphology2017,jiPhasefieldModelingPrecipitation2018, vaithyanathanMultiscaleModelingPrecipitation2004,kimFirstprinciplesPhasefieldModeling2017} there are only few works \cite{hauslerPrecipitationT1Phase2017} numerically investigating T$_1$ precipitates. The poor availability of relevant model constants and high aspect ratios of T$_1$ precipitates pose significant challenges from a simulation perspective. Additionally, stoichiometric multi-component compounds, such as T$_1$, are difficult to study using conventional phase-field approaches like the Kim-Kim-Suzuki (KKS) model \cite{kimPhasefieldModelBinary1999}. The approximation of the stoichiometric free energy using parabolic functions can lead to numerical instabilities and increase the uncertainty of the model predictions. To overcome this limitation, a novel phase-field modeling framework has been proposed by Ji and Chen \cite{jiPhasefieldModelStoichiometric2022} that uses the stoichiometric chemical potential and can thus overcome the limitation of KKS models for application to stoichiometric compounds.

In this work, we present a multi-component phase-field model for simulating the evolution of the T$_1$ phase. In section \ref{2} we analyze the crystallographic transformation of $\alpha$ to T$_1$ to investigate the expected elastic strains arising from lattice correspondence and group theory. Section \ref{3} presents the multiscale modeling strategy consisting of Density Functional Theory (DFT) for calculations of elastic moduli as well as the CALPHAD approach for calculation of thermodynamic equilibria. The obtained parameters are used in Cahn-Hilliard and Allen-Cahn evolution equations that model the kinetics of phase transformation and growth anisotropy. In section \ref{4} the results of one-, two- and three-dimensional simulations are presented that systematically highlight the capabilities of the model to simulate the precipitate growth accounting for different contributions of anisotropy with respect to the orientation variants predicted by lattice correspondence. We summarize our findings and conclusions in section \ref{5}.

\section{Crystallography of $\alpha$ to T$_1$ phase transformation}\label{2}

\begin{figure}[ht]
    \centering
    \includegraphics[width=0.5\textwidth]{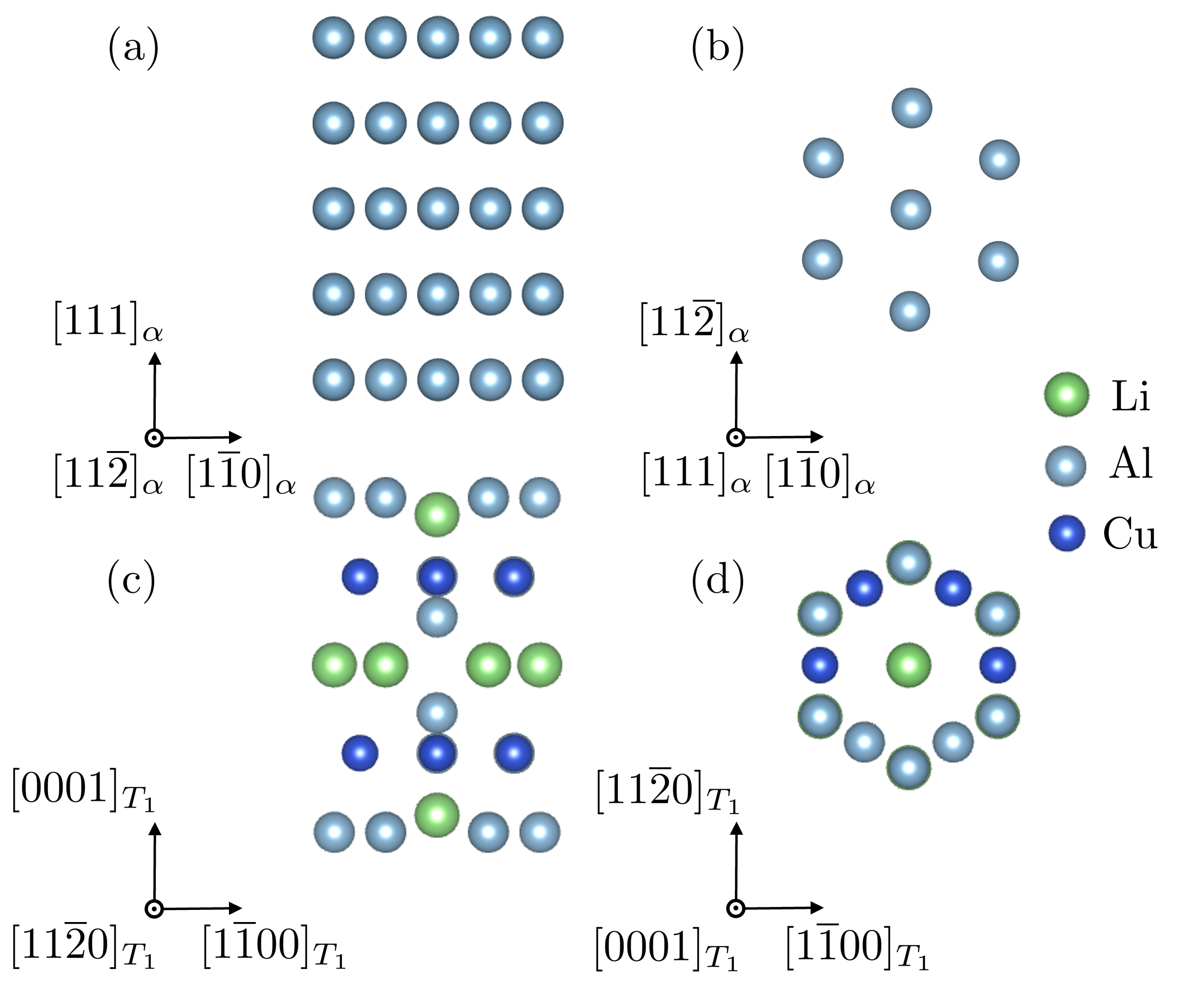}
    \caption{Structural model of $\alpha$-Al at the (a) $(11\overline{2})$  and  (b) $(111)$ viewplane and of T$_1$ at the (c) $(11\overline{2}0)$  and  (d) $(0001)$ viewplane.}
    \label{fig:LC}
\end{figure}

The transformation from the face-centered cubic (fcc) $\alpha$ structure to the hexagonal close-packed (hcp) T$_1$ phase is a process that involves changes in crystal symmetry and lattice parameters. The transformation can follow different pathways, each associated with changes in the point group symmetry dictated by the possible pathway degeneracies and broken symmetries. The parent phase in this transformation is the fcc $\alpha$ phase, characterized by a high degree of symmetry under the point group $m\overline{3}m$. The child phase is the hcp T$_1$ phase. This phase has lower symmetry than the fcc structure, typically described by the point group $6/mmm$ \cite{naFirstprinciplesCalculationsBulk2021}. Experimental findings \cite{niePhysicalMetallurgyLight2014} show that the orientation relationship can be described as $(111)_{\alpha} // (0001)_{\text{T}_1}$ and $[1\overline{1}0]_{\alpha} // [11\overline{2}0]_{\text{T}_1}$ for variant 1 as evident in Fig.~\ref{fig:LC}. The transformation involves symmetry breaking, which leads to various possible transformation pathways (TP) indicated as variant $p$. These variants are a result of the lattice correspondence (LC) and the symmetry operations retained during the transformation.

The number of variants resulting from the transformation of the $\alpha$ structure to the hcp T$_1$ phase, $N_{\alpha \rightarrow \text{T}_1}$, can be calculated using group theory \cite{gaoGroupTheoryDescription2016}, which takes into account the order of the point groups $\left|H_{}^{}\right|$ and the subgroup symmetries retained:
\begin{equation}
N_{\alpha \rightarrow \text{T}_1}=\frac{\left|H_{\alpha}^{\alpha}\right|}{\left|J^{\alpha / \text{T}_1}\right|}.
\end{equation}
Here, $\left|H_{\alpha}^{\alpha}\right|$ denotes the order of the parent matrix $\alpha$-Al point group in matrix coordinates, with $\left|J^{\alpha / \text{T}_1}\right|$ the order of the stabilizer subgroup that includes the symmetry operations preserved during the transformation, according to the LC. For Al, the point group is $m \overline{3} m$ which has an order of 48, whereas for the T$_1$ precipitate the point group is $6 / \mathrm{mmm}$, with an order of 24. Utilizing the previously identified LC, the subgroup linking $m \overline{3} m$ and $6 / \mathrm{mmm}$ is identified as $\overline{3} m$, making the stabilizer subgroup order $\left|J^{\alpha / \text{T}_1}\right|=12$. Consequently, $N_{\alpha \rightarrow \text{T}_1}$ equals 4. 

The stress-free transformation strains (SFTS) that occur during $\alpha$ to T$_1$ phase transformation can be calculated as follows:

\begin{equation}
\label{eq:strain}
\begin{aligned}
&\boldsymbol{\varepsilon}_p^0=\frac{\mathbf{T}_p^T\mathbf{T}_p-\mathbf{I}}{2},
\end{aligned}
\end{equation}

where $\mathbf{T}_p$ and $\mathbf{T}_p^T$ denotes the deformation gradient tensor of variant $p$ and its transpose, respectively. It maps the initial, undeformed state to its deformed state. $\mathbf{I}$ represents the identity matrix. The deformation gradient tensor can be assembled by solving an equation system with respect to three non-coplanar transformation vectors as follows:

\begin{equation}
\label{eq:DeformationGrad}
\begin{aligned}
&\mathbf{T}_p\mathbf{e}^{\alpha}_i = \mathbf{e}^{\text{T}_1}_i,
\end{aligned}
\end{equation}

where $\mathbf{e}^{\alpha}_i$ and $\mathbf{e}^{\text{T}_1}_i$ represent the non-coplanar vectors in $\alpha$ and $\text{T}_1$, respectively. From LC consideration, the following relations can be identified that yield the lowest strain magnitudes during bulk transformation:

\begin{align*}
4[111]_{\alpha} &\rightarrow [0001]_{\text{T}_1}, \\
\frac{1}{2}[11\overline{2}]_{\alpha} &\rightarrow \frac{1}{3}[11\overline{2}0]_{\text{T}_1}, \\
\frac{1}{2}[\overline{1}2\overline{1}]_{\alpha} &\rightarrow \frac{1}{3}[\overline{1}2\overline{1}0]_{\text{T}_1},
\end{align*}

which result in the following vector relations, considering the interplanar spacings, $d$, of the different directions:

\[\mathbf{e}^{\alpha}_1 = 4d_{111\alpha} 
\begin{bmatrix}
1 \\ 1 \\ 1
\end{bmatrix}_{\alpha} 
= \frac{4}{\sqrt{3}}a_{\alpha} 
\begin{bmatrix}
1 \\ 1 \\ 1
\end{bmatrix}_{\alpha} 
\quad \rightarrow \quad \mathbf{e}^{\text{T}_1}_1 = d_{0001\text{T}_1}
\begin{bmatrix}
1 \\1 \\ 1
\end{bmatrix}_{\alpha}
= c_{\text{T}_1}
\begin{bmatrix}
1 \\1 \\ 1
\end{bmatrix}_{\alpha}, \]

\[\mathbf{e}^{\alpha}_2 = \frac{1}{2}d_{11\overline{2}\alpha}
\begin{bmatrix}
1 \\1 \\ \overline{2}
\end{bmatrix}_{\alpha}
=\frac{1}{2\sqrt{6}}a_{\alpha}
\begin{bmatrix}
1 \\1 \\ \overline{2}
\end{bmatrix}_{\alpha}
\quad \rightarrow \quad \mathbf{e}^{\text{T}_1}_2 = \frac{1}{3} d_{11\overline{2}0\text{T}_1}
\begin{bmatrix}
1 \\ 1 \\ \overline{2}
\end{bmatrix}_{\alpha}
= \frac{1}{6} a_{\text{T}_1}
\begin{bmatrix}
1 \\ 1 \\ \overline{2}
\end{bmatrix}_{\alpha},\]

\[\mathbf{e}^{\alpha}_3 = \frac{1}{2}d_{\overline{1}2\overline{1}\alpha}
\begin{bmatrix}
\overline{1} \\ 2 \\ \overline{1}
\end{bmatrix}_{\alpha}
= \frac{1}{2\sqrt{6}}a_{\alpha}
\begin{bmatrix}
\overline{1} \\ 2 \\ \overline{1}
\end{bmatrix}_{\alpha}
\quad \rightarrow \quad \mathbf{e}^{\text{T}_1}_3 = \frac{1}{3} d_{\overline{1}2\overline{1}0\text{T}_1}
\begin{bmatrix}
\overline{1} \\ 2 \\ \overline{1}
\end{bmatrix}_{\alpha}
= \frac{1}{6} a_{\text{T}_1}
\begin{bmatrix}
\overline{1} \\ 2 \\ \overline{1}
\end{bmatrix}_{\alpha},\]

where $a_{\alpha} = 4.049$~\r{A}, $a_{\text{T}_1} = 4.954$~\r{A} and $c_{\text{T}_1} = 9.327$~\r{A} denote the lattice parameters of $\alpha$ and $\text{T}_1$, respectively. While this transformation relation holds true for bulk T$_1$ structures, we emphasize the difference with respect to the experimentally observed flat structures that consists of only a few atomic layers and its consequences for the effective transformation strains. According to Häusler et al. \cite{hauslerThickeningT1Precipitates2018}, thickening occurs as alternating stacking of Cu and Li-rich layers. They can be classified into 4 different types according to their thickness: 3 layers (0.505 nm), 7 layers (1.471 nm), 11 layers (2.437 nm) and 15 layers (3.403 nm). To determine the deformation gradient components and the corresponding strains acting on $(111)_{\alpha} // (0001)_{\text{T}_1}$, we identify the number of $(111)_{\alpha}$ layers which yield the lowest strain magnitudes: 2 layers (8.33\%), 7 layers (-9.6\%), 11 layers (-5.09\%) and 15 layers (-2.91\%). For simplicity, we assume a strain of -5.09\%, corresponding to a thickness of 2.437 nm, which closely aligns with the average thickness observed after extended aging treatment in the study by Häusler et al. \cite{hauslerThickeningT1Precipitates2018}. The resulting deformation gradient tensor and the effective SFTS are listed in Table \ref{tbl:Eigenstrains}.


\begin{table*}[ht]
\caption{Deformation gradient and SFTS for the identified transformation pathways}
\centering
\begin{tabularx}{\columnwidth}{>{\centering\arraybackslash}p{0.1\columnwidth}>{\centering\arraybackslash}p{0.2\columnwidth}>{\centering\arraybackslash}p{0.3\columnwidth}>{\centering\arraybackslash}p{0.3\columnwidth}}
\toprule
Variant & Orientation relationship & Deformation gradient& SFTS \\
\midrule
1 & \begin{tabular}{@{}c@{}}$(111)_{\alpha} // (0001)_{\text{T}_1}$ \\ $[11\overline{2}]_{\alpha} // [11\overline{2}0]_{\text{T}_1}$\end{tabular} &
$\begin{pmatrix}
0.9819 & -0.0171 & -0.0171 \\
-0.0171 & 0.9819 & -0.0171 \\
-0.0171 & -0.0171 & 0.9819 \\
\end{pmatrix}$ &
$\begin{pmatrix}
-0.0176 & -0.0166 & -0.0166 \\
-0.0166 & -0.0176 & -0.0166 \\
-0.0166 & -0.0166 & -0.0176 \\
\end{pmatrix}$ \\[20pt]
2  & \begin{tabular}{@{}c@{}}$(\overline{1}11)_{\alpha} // (0001)_{\text{T}_1}$ \\ $[\overline{1}1\overline{2} ]_{\alpha} // [11\overline{2}0]_{\text{T}_1}$\end{tabular} &
$\begin{pmatrix}
0.9819 & 0.0171 & 0.0171 \\
0.0171 & 0.9819 & -0.0171 \\
0.0171 & -0.0171 & 0.9819 \\
\end{pmatrix}$ &
$\begin{pmatrix}
-0.0176 & 0.0166 & 0.0166 \\
0.0166 & -0.0176 & -0.0166 \\
0.0166 & -0.0166 & -0.0176 \\
\end{pmatrix}$ \\[20pt]
3 & \begin{tabular}{@{}c@{}}$(1\overline{1}1)_{\alpha} // (0001)_{\text{T}_1}$ \\ $[1\overline{1}\overline{2} ]_{\alpha} // [11\overline{2}0]_{\text{T}_1}$\end{tabular} &
$\begin{pmatrix}
0.9819 & 0.0171 & -0.0171 \\
0.0171 & 0.9819 & 0.0171 \\
-0.0171 & 0.0171 & 0.9819 \\
\end{pmatrix}$ &
$\begin{pmatrix}
-0.0176 & 0.0166 & -0.0166 \\
0.0166 & -0.0176 & 0.0166 \\
-0.0166 & 0.0166 & -0.0176 \\
\end{pmatrix}$ \\[20pt]
4 & \begin{tabular}{@{}c@{}}$(11\overline{1})_{\alpha} // (0001)_{\text{T}_1}$ \\ $[112]_{\alpha} // [11\overline{2}0]_{\text{T}_1}$\end{tabular} &
$\begin{pmatrix}
0.9819 & -0.0171 & 0.0171 \\
-0.0171 & 0.9819 & 0.0171 \\
0.0171 & 0.0171 & 0.9819 \\
\end{pmatrix}$ &
$\begin{pmatrix}
-0.0176 & -0.0166 & 0.0166 \\
-0.0166 & -0.0176 & 0.0166 \\
0.0166 & 0.0166 & -0.0176 \\
\end{pmatrix}$ \\
\bottomrule
\end{tabularx}
\label{tbl:Eigenstrains}
\end{table*}

\section{Methodology} \label{3}
This study employs a multiscale approach to simulate T$_1$ phase precipitation. The phase-field model is central to this research, with parameters obtained from first-principle calculations and CALPHAD databases. Despite the disagreement in the literature regarding the structural configuration and stoichiometry of T$_1$, we assume a stoichiometry of Al$_2$CuLi as suggested by van Smaalen \cite{vansmaalenRefinementCrystalStructure1990} because reliable thermochemical data of this structure is available in CALPHAD databases \cite{saunders1998cost}.
\subsection{Phase-field model} \label{PF}
We consider a homogeneous Al-Cu-Li alloy where the constituents Al, Cu and Li are in solid solution. The stoichiometric reversible reaction to precipitate T$_1$ from solid-solution can be described as:
\begin{equation}
\label{eq:reaction}
\begin{aligned}
& v_{\text{Al}}\text{Al} + v_{\text{Cu}}\text{Cu} +  v_{\text{Li}}\text{Li} \rightleftharpoons  \text{Al}_{v_{\text{Al}}}\text{Cu}_{v_{\text{Cu}}}\text{Li}_{v_{\text{Li}}},
\end{aligned}
\end{equation}

where $v_{i}$ denotes the stoichiometric coefficient of the constituent element $i$. The coefficients are normalized, i.e. $\sum^i v_i = 1$. In the stoichiometric compound $\text{Al}_2\text{Cu}\text{Li}$, the coefficients hold $v_{\text{Al}} = 0.5$, $v_{\text{Cu}} = 0.25$, and $v_{\text{Li}} = 0.25$. We introduce a non-conserved order parameter variable, $\eta_p$, which tracks the extent of the stoichiometric reaction described in Eq.~(\ref{eq:reaction}) for variant $p$ that identifies the crystallographic variant from LC. Here, $\eta_p = 0$ indicates the matrix phase and  $\eta_p = 1$ refers to the precipitate variant phase $p$. The total composition is determined as:
\begin{equation}
\label{eq:total_comp}
\begin{aligned}
c_i^{\text{tot}} = c_i + v_{i}\sum_{p=1}^4 \eta_p,
\end{aligned}
\end{equation}

where $c_i^{\text{tot}}$ and $c_i$ denote the total composition of element $i$ and the composition in solute, respectively. The total free energy of the system, $F$, is decomposed into three terms as:

\begin{equation}
\label{eq:free-energy_functional}
\begin{aligned}
F=\int\left[f_{\text{bulk}}+f_{\text{int}}+f_{\text{el}}\right] d V,
\end{aligned}
\end{equation}

where $f_{\text{bulk}}$ is the bulk free energy density, $f_{\text{int}}$ is the interfacial free energy density, and $f_{\text{el}}$ is the elastic free energy density.

\subsubsection{Chemical free energy}\label{chemical}
Ji and Chen \cite{jiPhasefieldModelStoichiometric2022} have proposed a comprehensive approach to describe the bulk free energy density of a multiphase system containing stoichiometric compounds. The formulation of the bulk free energy density interpolates between the contribution of stoichiometric compound and solid solution matrix, which is expressed as: 
\begin{equation}
\label{eq:bulk}
\begin{aligned}
f_{\text{bulk}} = \frac{1}{V_m} \left[ H(\boldsymbol{\eta}) \mu^{\text{T}_{\text{1}}} +  \left[1 - H(\boldsymbol{\eta}) \right] \mu^{\alpha} (\mathbf{c}) \right],
\end{aligned}
\end{equation}

where \(V_m\) denotes the molar volume, and $H(\boldsymbol{\eta})$ is a multi-order-parameter interpolation function that reads \(H(\boldsymbol{\eta}) = \sum_{p=1}^4 h(\eta_p)\), and $h(\eta_p$) is Wang's interpolation function \cite{wangThermodynamicallyconsistentPhasefieldModels1993} of the variant $p$ given as:
\begin{equation}
\begin{aligned}
h(\eta_p) = 6\eta_p^5 - 15\eta_p^4 + 10 \eta_p^3.
\end{aligned}
\end{equation}

\begin{figure*}[ht]
    \centering
    \includegraphics[width=0.7\textwidth]{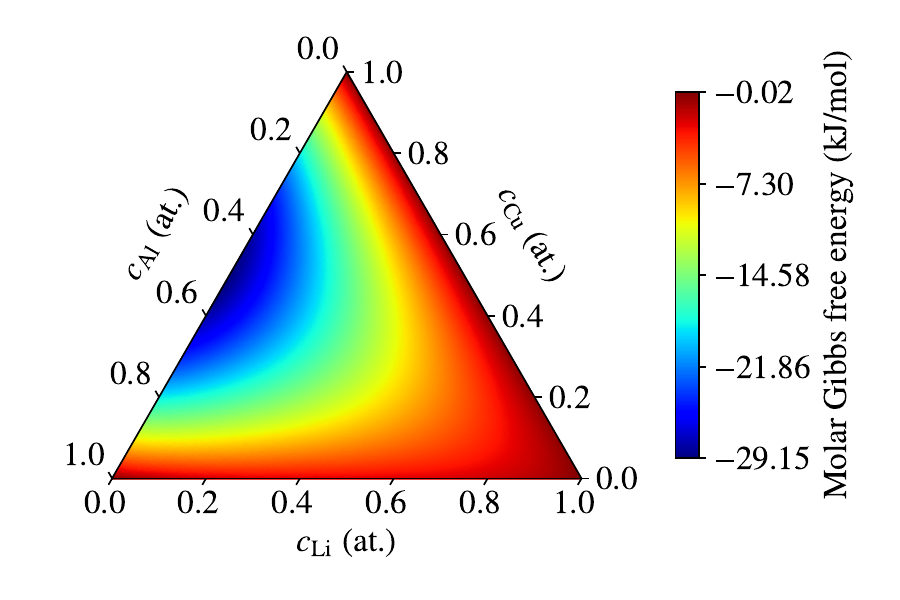}
    \caption{Ternary chemical potential/molar Gibbs free energy diagram for the $\alpha$ phase in the Al-Cu-Li system at 155~$^{\circ}\mathrm{C}$.}
    \label{fig:Gibbs}
\end{figure*}

The full derivation of Eq.~(\ref{eq:bulk}) is provided in Appendix~\ref{app1}. The term \(\mu^{\text{T}_{\text{1}}}\) signifies the stoichiometric chemical potential of the T$_1$ phase. \(\mu^{\alpha} (\mathbf{c})\) represents the chemical potential or molar Gibbs free energy \cite{chenChemicalPotentialGibbs2019} of the $\alpha$-phase as a function of the composition in solute which can be described using the regular solution model for the Al-Cu-Li alloy system, neglecting magnetic effects:
\begin{equation}
\mu^{\alpha} \left(\mathbf{c}, T\right) =  \sum_i c_i\mu^{\alpha}_i=\sum_i c_i \left[\mu^{0}_i +\sum_{j>i} L_{i j}^{\alpha} c_j+R T \ln c_i \right],
\label{eq:reg-sol}
\end{equation}

where $\mu^{\alpha}_i$ is the chemical potential of the solute elements in the $\alpha$-phase, $\mu^{0}_i$ is the chemical potential of the pure components, $ L_{i j}^{\alpha}$ is the binary interaction coefficient in the excess chemical potential term, $R$ is the gas constant and $T$ is the temperature. ${ }^0 \mu^{\alpha}_i$ is described by a polynomial as function of temperature \cite{saunders1998cost}, which is expressed as:
\begin{equation}
\begin{aligned}
\mu^{0}_i = A^{0}_i + B^{0}_i T + C^{0}_i T^{2} + D^{0}_i T^{3} + E^{0}_iT \ln T + F^{0}_iT^{-1},
\end{aligned}
\end{equation}

where $\{A^{0}_i, B^{0}_i, ...,F^{0}_i\} $ denote the set of fitted coefficients in the chemical potential polynomial of the pure element $i$. The corresponding values used in this work are listed in Table~\ref{tbl:Gibbs_alpha_coeff} of Appendix~\ref{app4}. The deviation from the ideal solution is expressed by the excess terms that capture the non-ideal interactions between different species in the alloy and is described using \(L_{i j}^{\alpha}\). A possible ternary interaction function is neglected in the current formalism. The binary interaction functions take the form of Riedlich-Kister polynomials \cite{redlichAlgebraicRepresentationThermodynamic1948} in the order of $n$ and are defined as: 

\begin{equation}
\begin{aligned}
L_{i j}^{\alpha} = \sum_{k}^{n} {}^{k}L_{i j}^{\alpha} \left[c_{i} - c_{j}\right]^{k},
\end{aligned}
\end{equation}

where ${}^{k}L_{i j}^{\alpha}$ represent the temperature-dependent polynomial coefficients, which are listed in Table~\ref{tbl:binary_interaction} of Appendix~\ref{app4}. The formulation of chemical potentials in the regular solution model, as depicted in Eq.~(\ref{eq:reg-sol}), allows for the analysis of energetics across varying temperatures and compositions. Fig.~\ref{fig:Gibbs} illustrates the values of the chemical potential of the $\alpha$-phase at a specific temperature of 155\,$^\circ$C.

The values for the interaction coefficients are listed in Appendix~\ref{app4}, respectively. The chemical potential of the stoichiometric compound is described as a temperature-dependent function:

\begin{equation}
\begin{aligned}
\mu^{\text{T}_{\text{1}}} = A^{\text{T}_{\text{1}}}+ B^{\text{T}_{\text{1}}} T +v_{\text{Al}} \mu^{0}_\text{Al} + v_{\text{Cu}} \mu^{0}_\text{Cu} + v_{\text{Li}} \mu^{0}_\text{Li},
\end{aligned}
\end{equation}

where $A^{\text{T}_{\text{1}}} = -24560.0$ and $B^{\text{T}_{\text{1}}}=6.0$ denote the polynomial coefficients \cite{saunders1998cost}. The driving force for phase transformation can be determined by taking into account the differences between the chemical potential of the stoichiometric compound and the weighted chemical potential of the solute elements as described via:

\begin{equation}
\label{eq:driving-force}
\begin{aligned}
\Delta \mu^{r} (\mathbf{c}) = \mu^{\text{T}_{\text{1}}} - \frac{1}{2} \mu_{\text{Al}} - \frac{1}{4} \mu_{\text{Cu}} - \frac{1}{4} \mu_{\text{Li}}.
\end{aligned}
\end{equation}

\begin{figure*}[ht]
    \centering
    \includegraphics[width=0.7\textwidth]{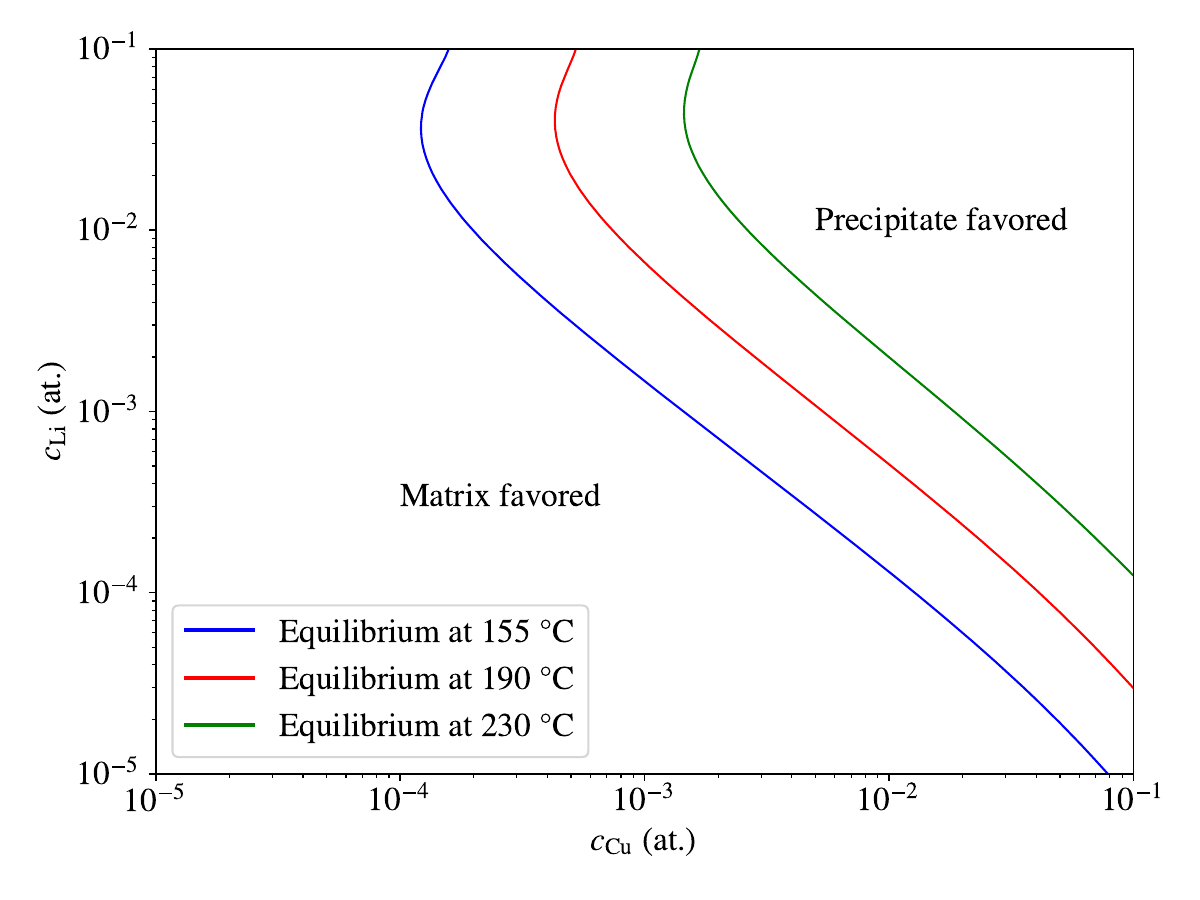}
    \caption{Chemical potential driving force according to Eq.~(\ref{eq:driving-force}) and the equilibrium lines at 155~$^{\circ}\mathrm{C}$, 190~$^{\circ}\mathrm{C}$ and 230~$^{\circ}\mathrm{C}$, which correspond to common temperatures for the aging heat treatment of Al-Cu-Li alloys. }
    \label{fig:DF}
\end{figure*}

The evaluation of Eq.~(\ref{eq:driving-force}) allows to define equilibrium states across the compositional spectrum in ternary space. As shown in Fig.~\ref{fig:DF}, the driving force defines areas where the reaction favors matrix formation and other areas where precipitate formation is favored. The temperature-dependent equilibrium state is defined at compositions where Eq.~(\ref{eq:driving-force}) takes the value of 0.

\subsubsection{Interfacial free energy}\label{interfacial}

The interfacial free energy is decomposed into a double-well free energy contribution and a gradient-dependent term as:
\begin{equation}
\label{eq:IE}
\begin{aligned}
f_{\text{int}} = w g(\boldsymbol{\eta}) +  \sum_{p=1}^4\frac{1}{2} \boldsymbol{\kappa}_{p} \cdot \left[\nabla \eta_p \otimes\nabla \eta_p \right],
\end{aligned}
\end{equation}

where $w$ is the double-well height, $\cdot$ denotes a scalar product, $\otimes$ represents the dyadic product, and $g(\boldsymbol{\eta})$ is an interpolation function, which is defined as: 

\begin{equation}
\label{eq:DW}
\begin{aligned}
g(\boldsymbol{\eta}) = \sum_{p=1}^4 \eta_p^2\left[1-\eta_p\right]^2+5 \sum_{p, q>p}^4 \eta_p^2 \eta_q^2.
\end{aligned}
\end{equation}

In Eq.~(\ref{eq:IE}), the gradient energy coefficient $\boldsymbol{\kappa}_{p}$ is a second-order tensor that captures the interfacial energy anisotropy of the T$_1$/$\alpha$ interfaces. In a coordinate system where the principal directions align with the normal directions of the basal and periphery planes in the T$_1$ structure, the gradient energy coefficient tensor can be expressed using the following diagonal matrix:

\begin{equation}
\label{eq:grcoeff}
\boldsymbol{\kappa} =  \begin{pmatrix}
\kappa_{11} & 0 & 0 \\
0 & \kappa_{22} & 0 \\
0 & 0 & \kappa_{33} \\
\end{pmatrix},
\end{equation}

where $\kappa_{22}$ refers to the gradient energy coefficient of the $(111)_{\alpha} // (0001)_{\text{T}_1}$ interface and $\kappa_{11} = \kappa_{33}$ denote the coefficients of the $(1\overline{1}0)_{\alpha} // (11\overline{2}0)_{\text{T}_1}$ interfaces. The gradient energy coefficients $\kappa_{ii}$ and the double-well energy height are related to the interfacial energy $\gamma$ and interfacial thickness $\lambda$ in the following manner \cite{jiPhasefieldModelStoichiometric2022}:

\begin{equation}
\begin{aligned}
\kappa = \frac{3}{2}\gamma\lambda,\\
w = 12 \frac{\gamma}{\lambda}.
\end{aligned}
\end{equation}

The second-order tensor in Eq.~(\ref{eq:grcoeff}) has to be rotated to the reference coordinate system of the $\alpha$-Al so that the transformed principal directions align as follows: $[\overline{1} 1 0]\rightarrow [1 0 0] $, $[1 1 1]\rightarrow [0 1 0] $, $[1 1 \overline{2}]\rightarrow [0 0 1]$. The rotation of $\boldsymbol{\kappa}$ to $\boldsymbol{\kappa}_p$ can be mathematically described using the following operation:

\begin{equation}
\boldsymbol{\kappa}_p = \boldsymbol{R}_p \boldsymbol{\kappa} \boldsymbol{R}_p^T,
\end{equation}

where $\boldsymbol{R}_p$ and $\boldsymbol{R}_p^T$ denote the rotation matrix and its transpose, respectively. $\boldsymbol{R}_p$ for variant $p=1$ is constructed by using the normalized orthogonal vectors that provide the new principal directions. The rotation of the gradient energy coefficient matrix for the other variants can be performed similarly in accordance to the LC. The values for the tensors $\boldsymbol{R}_p$ of all the variants are listed in Appendix~\ref{app4}.

\subsubsection{Elastic strain energy}\label{elasticity}
The contribution of the elastic strain energy can be determined using microelasticity theory \cite{khachaturyanTheoryStructuralTransformation1983}. Several methods have been developed \cite{jinThreedimensionalPhaseField2001, michelEffectivePropertiesComposite1999, moulinecNumericalMethodComputing1998, gururajanPhaseFieldStudy2007} to solve the problem numerically. To evaluate the elastic strain energy in phase-field models computationally, one must solve coupled differential equations that govern the phase-field evolution and mechanical equilibrium. The elastic strain energy density, \(f_{\text{el}}\), is a measure of the energy stored in a material due to elastic deformation and is expressed as:
\begin{equation}
f_{\text{el}} = \frac{1}{2} \boldsymbol{\sigma}^{\text{el}}  \cdot \boldsymbol{\varepsilon}^{\text{el}},
\end{equation}
where \( \boldsymbol{\sigma}^{\text{el}} \) represents the elastic stress tensor, and \( \boldsymbol{\varepsilon}^{\text{el}} \) denotes the elastic strain tensor. Assuming linear elasticity, the elastic stress tensor \( \boldsymbol{\sigma}^{\text{el}} \) is related to the elastic strain \( \boldsymbol{\varepsilon}^{\text{el}} \) through the fourth-order elastic stiffness tensor \(\mathcal{C}\), in the following manner:
\begin{equation}
\boldsymbol{\sigma}^{\text{el}} = \mathcal{C} \boldsymbol{\varepsilon}^{\text{el}}.
\end{equation}

The total strain \( \boldsymbol{\varepsilon} \) in the material can be decomposed into the elastic part \( \boldsymbol{\varepsilon}^{\text{el}} \) and the SFTS, \( \boldsymbol{\varepsilon}^0 \), that captures the effects of the TP as determined in Section \ref{2}:
\begin{equation}
\boldsymbol{\varepsilon}^{\text{el}} = \boldsymbol{\varepsilon} - \boldsymbol{\varepsilon}^0.
\end{equation}

The total SFTS interpolates between the variant-specific SFTS $\boldsymbol{\varepsilon}_{p}^{0}$ as follows:
\begin{equation}
\boldsymbol{\varepsilon}^0(\boldsymbol{\eta})=\sum_{p=1}^4 h(\eta_p) \boldsymbol{\varepsilon}_{p}^{0}.
\end{equation}

The phase dependent elastic stiffness tensor can be described as:

\begin{equation}
\label{eq:Int_ETensor}
\mathcal{C}\left(\boldsymbol{\eta}\right)= \left[ 1-H\left(\boldsymbol{\eta}\right)\right] \mathcal{C}^{\alpha} + H\left(\boldsymbol{\eta}\right) \mathcal{C}^{\text{T$_1$}},
\end{equation}

where $\mathcal{C}^{\alpha}$ and $\mathcal{C}^{\text{T}_1}$ are the elasticity tensors of $\alpha$-Al and ${\text{T}_1}$, respectively. To acquire the elastic strains to evaluate the contribution of the elastic strain energy, we solve the mechanical equilibrium that can be stated as:

\begin{equation}
\text{div}(\boldsymbol{\sigma}^{\text{el}}) = \mathbf{0},
\end{equation}

with $\text{div}(\cdot)$ representing the divergence operator. The mechanical equilibrium is solved by employing the spectral method \cite{moulinecNumericalMethodComputing1998} and transforming the equilibrium equations into the frequency domain. The acoustic tensor $K^0$ is defined for a reference homogeneous material as: 

\begin{equation}
K^0(\xi) = \mathcal{C}^0 \left[\xi \otimes \xi \right],
\end{equation}

where $\boldsymbol{\xi}$ denotes the frequency space vector. It has been shown that choosing $\mathcal{C}^0$ to be the average of the stiffness of matrix and precipitate phase leads to optimal convergence behaviour \cite{gururajanPhaseFieldStudy2007}. The Green's function, $\hat{\Gamma}$, can be assembled using the inverse of the acoustic tensor $N^0 = [K^0]^{-1}$. By utilizing Green's functions and the spectral method, we can update the mechanical fields and ensure equilibrium as summarized in Algorithm~(\ref{alg:FFT_Update}). The stress tensors are updated using Hooke's law, and the acting elastic strains are computed accordingly. In terms of index notation, the Green's function \( \hat{\Gamma}_{khij} \) is expressed as:

\begin{equation}
\begin{aligned}
\hat{\Gamma}_{k h i j}= & \frac{1}{4}\left[N_{h i}^0(\boldsymbol{\xi}) \xi_j \xi_k+N_{k i}^0(\boldsymbol{\xi}) \xi_j \xi_h+N_{h j}^0(\boldsymbol{\xi}) \xi_i \xi_k +N_{k j}^0(\boldsymbol{\xi}) \xi_i \xi_h\right].
\end{aligned}
\end{equation}

\begin{algorithm}
\caption{FFT-based stress-strain field update}\label{alg:FFT_Update}
\begin{algorithmic}
\Require Initial stress $\boldsymbol{\sigma}^0$ and strain $\boldsymbol{\varepsilon}^0$ fields are given.
\Ensure Updated fields $\boldsymbol{\sigma}^{i+1}$ and $\boldsymbol{\varepsilon}^{i+1}$ after convergence.
\State $\hat{\boldsymbol{\sigma}}^i \gets \text{FFT}(\boldsymbol{\sigma}^i)$
\State $\hat{\boldsymbol{\varepsilon}}^i \gets \text{FFT}(\boldsymbol{\varepsilon}^i)$
\State $i \gets 0$ \Comment{Initialize iteration counter}
\Repeat
    \State $\hat{\boldsymbol{\varepsilon}}^{i+1}(\boldsymbol{\xi}) \gets \hat{\boldsymbol{\varepsilon}}^i(\boldsymbol{\xi}) - \hat{\Gamma}(\boldsymbol{\xi}) \hat{\boldsymbol{\sigma}}^i(\boldsymbol{\xi})$
    \State $\boldsymbol{\varepsilon}^{i+1}(\mathbf{x}) \gets \text{FFT}^{-1}(\hat{\boldsymbol{\varepsilon}}^{i+1}(\boldsymbol{\xi}))$
    \State $\boldsymbol{\sigma}^{i+1}(\mathbf{x}) \gets \mathcal{C}(\mathbf{x}) \boldsymbol{\varepsilon}^{i+1}(\mathbf{x})$
    \State $\hat{\boldsymbol{\sigma}}^{i+1} \gets \text{FFT}(\boldsymbol{\sigma}^{i+1})$
    \State $i \gets i + 1$
    \State Check convergence $\int_V\| \boldsymbol{\sigma}^{i+1}(\mathbf{x}) - \boldsymbol{\sigma}^i(\mathbf{x}) \|^2dV < \text{tol}$
\Until{converged}
\end{algorithmic}
\end{algorithm}

Elastic energy contributions are detailed through variational derivative of elastic strain energy with respect to the order parameter and components:
\begin{equation}
\frac{\partial f_{e l}}{\partial \eta_p}=-\mathcal{C} \left[\boldsymbol{\varepsilon}-\boldsymbol{\varepsilon}^0\right] \cdot \frac{\partial h(\eta_p)}{\partial \eta_p} \boldsymbol{\varepsilon}_{p}^{0}.
\end{equation}


\subsubsection{Evolution equations}\label{evol}
Phase evolution and diffusion is driven by the minimization of the total free energy functional in Eq.~(\ref{eq:free-energy_functional}). The Allen-Cahn equation determines the evolution of the reaction as stated in Eq.~(\ref{eq:reaction}): 

\begin{equation}
\frac{\partial \eta_p}{\partial t}=-L_{\eta_p} \frac{\partial F}{\partial \eta_p} =-L_{\eta_p}\left[\frac{\partial h(\eta_p)}{\partial \eta_p} V_m^{-1} \Delta \mu^r\left(\mathbf{c}\right)+w \frac{\partial g(\boldsymbol{\eta})}{\partial \eta_p}-\boldsymbol{\kappa}_p \cdot \nabla^2 \eta_p+\frac{\partial f_{e l}}{\partial \eta_p}\right],
\end{equation}

where $L_{\eta_p}$ denotes the reaction rate of variant $p$. 
The Cahn-Hilliard evolution equations for Cu and Li are formulated to capture diffusion-driven transformations. These include terms for mobility and chemical potential gradients on the concentration profiles \cite{jiPhasefieldModelStoichiometric2022}:

\begin{equation}
\label{eq:CH}
\begin{aligned}
\frac{\partial c_\text{Cu}}{\partial t} &= \nabla \cdot \left[ M_\text{Cu}(\boldsymbol{\eta})\,\nabla\left[V_m^{-1} \left[ \mu_\text{Cu}(\mathbf{c}) - \mu_\text{Al}(\mathbf{c}) \right] \right]\right] - \frac{\partial \left[ H(\boldsymbol{\eta}) \left[ v_{\text{Cu}} - c_\text{Cu} \right] \right]}{\partial t}, \\
\frac{\partial c_\text{Li}}{\partial t} &= \nabla \cdot \left[ M_\text{Li}(\boldsymbol{\eta})\,\nabla\left[ V_m^{-1} \left[ \mu_\text{Li}(\mathbf{c}) - \mu_\text{Al}(\mathbf{c}) \right] \right] \right] - \frac{\partial \left[ H(\boldsymbol{\eta}) \left[ v_{\text{Li}} - c_\text{Li} \right] \right]}{\partial t},
\end{aligned}
\end{equation}

where the terms  $\mu_\text{Cu}\left(\mathbf{c}\right)-\mu_\text{Al}\left(\mathbf{c}\right)$ and $\mu_\text{Li}\left(\mathbf{c}\right)-\mu_\text{Al}\left(\mathbf{c}\right)$ are known as the diffusion potentials of Cu and Li, respectively. It is emphasized that the source terms on the right-hand side in Eq.~(\ref{eq:CH}) originate because we solve for $c_i$ instead of $c_i^{\text{tot}}$. The corresponding derivation is provided in Appendix~\ref{app2}. The chemical mobilities $M_\text{Cu}(\boldsymbol{\eta})$ and $M_\text{Li}(\boldsymbol{\eta})$ capture the phase-dependent diffusion kinetics and are defined as: 

\begin{equation}
\label{eq:Mobility}
\begin{aligned}
M_\text{Cu}(\boldsymbol{\eta}) = \frac{V_m D_\text{Cu}(\boldsymbol{\eta})}{\frac{\partial \mu_\text{Cu}\left(\mathbf{c}\right)}{\partial c_\text{Cu}}-\frac{\partial \mu_\text{Al}\left(\mathbf{c}\right)}{\partial c_\text{Cu}}}, \\
M_\text{Li}(\boldsymbol{\eta}) = \frac{V_m D_\text{Li}(\boldsymbol{\eta})}{\frac{\partial \mu_\text{Li}\left(\mathbf{c}\right)}{\partial c_\text{Li}}-\frac{\partial \mu_\text{Al}\left(\mathbf{c}\right)}{\partial c_\text{Li}}}.
\end{aligned}
\end{equation}

\begin{figure*}[ht]
    \centering
    \includegraphics[width=1\textwidth]{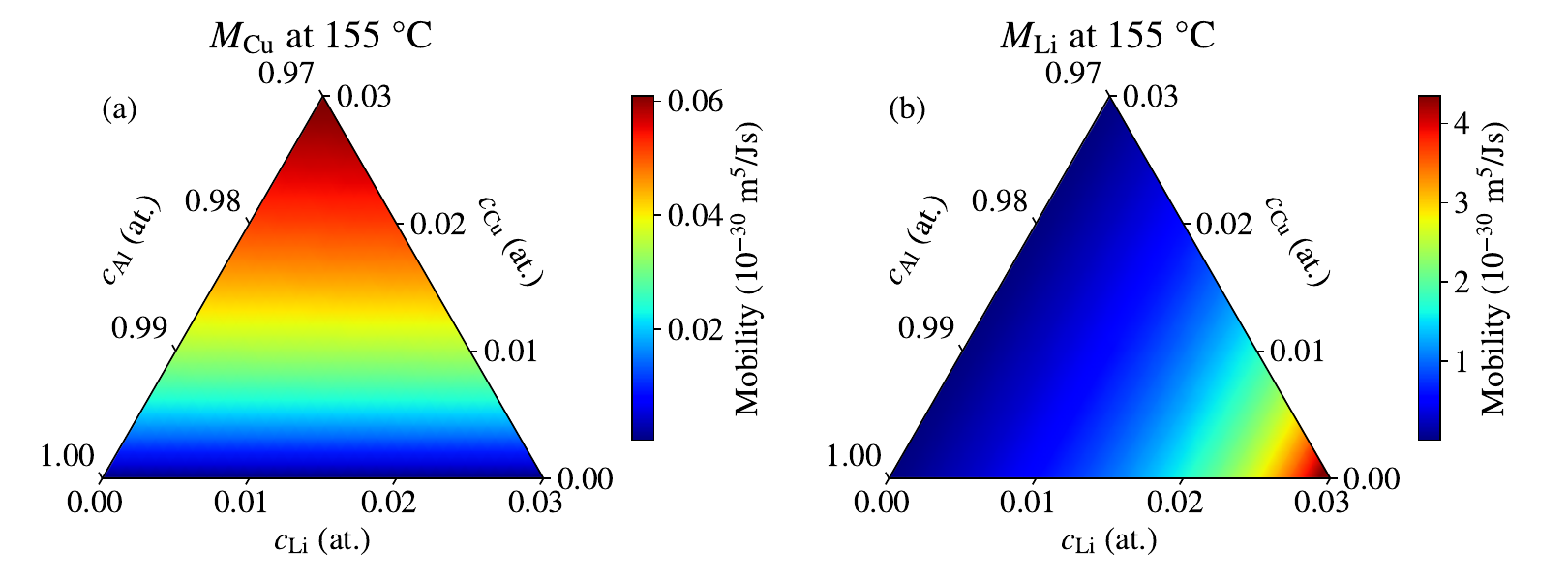}
    \caption{Chemical mobility diagrams for the interdiffusion of (a) Cu (\( M_\text{Cu} \)) and (b) Li (\( M_\text{Li} \)) in the solution matrix at 155~$^{\circ}\mathrm{C}$.}
    \label{fig:Mobility}
\end{figure*}

In Eq.~(\ref{eq:Mobility}) the diffusivity, $D_i(\boldsymbol{\eta})$, is divided by the derivative of the diffusion potential to ensure constant diffusivity across the phases. The diffusion coefficient of a species $i$ can be interpolated between the coefficients of solution matrix and stoichiometric compound as follows:

\begin{equation}
\begin{aligned}
D_i(\boldsymbol{\eta}) = \left[ 1- H(\boldsymbol{\eta}) \right] D_i^{\alpha} + H(\boldsymbol{\eta}) D_i^{\text{T}_1},
\end{aligned}
\end{equation}

where $D_i^{\alpha}$ and $D_i^{\text{T}_1}$ are the diffusion coefficients of species $i$ in $\alpha$-Al and $\text{T}_1$, respectively. In this work, we assume that the diffusion coefficients are the same for $\alpha$ and $\text{T}_1$. The diffusion coefficient of element $i$ at a given temperature $T$ reads as:

\begin{equation}
\begin{aligned}
D^{\alpha}_i = {}^{0}D^{\alpha}_i \exp \left[ -\frac{Q^{\alpha}_{i}}{RT} \right],
\end{aligned}
\end{equation}

where ${}^{0}D^{\alpha}_i$ denotes the pre-exponential diffusion term, $Q^{\alpha}_{i}$ is the energy barrier to activate diffusion and $R$ is the universal gas constant. The chemical mobilities are calculated for the Al-rich corner as shown in Fig.~\ref{fig:Mobility}.

We formulate the stoichiometric linear reaction rate, $L_{\eta_p}$, in the Allen-Cahn equation as an anisotropic function of the interface normal angle $\theta_p$ as follows: \cite{huThermodynamicDescriptionGrowth2007,salvalaglioFacetingEquilibriumMetastable2015,jiPhasefieldModelingPrecipitation2018}
\begin{equation}
\label{eq:L_Anisotropy}
L_{\eta_p}(\theta_p)=\frac{L_{\eta_p}^0}{1+\beta}\left\{\begin{array}{c}
1+\frac{\beta}{\sin \phi_0}+\frac{\beta \cos \phi_0}{\sin \phi_0} \sin \theta_p,-\frac{\pi}{2} \leq \theta_p \leq-\frac{\pi}{2}+\phi_0 \\
1+\beta \cos \theta_p,-\frac{\pi}{2}+\phi_0 \leq \theta_p \leq \frac{\pi}{2}-\phi_0 \\
1+\frac{\beta}{\sin \phi_0}-\frac{\beta \cos \phi_0}{\sin \phi_0} \sin \theta_p, \frac{\pi}{2}-\phi_0 \leq \theta_p \leq \frac{\pi}{2}
\end{array}\right. .
\end{equation}

Here, $L_{\eta_p}^0$ describes the stoichiometric reaction rate on the diffusion-controlled periphery interface and $\beta$ defines the anisotropy with respect to the interface-controlled basal interface. In general, wherever the stoichiometric reaction rate is high, the particle evolution can be considered diffusion-controlled. On the opposite side, when the stoichiometric reaction rate is low, the precipitate growth is controlled by the interfacial energy. $\phi_0$ is a small regularization angle that controls the transition smoothness in Eq.~(\ref{eq:L_Anisotropy}). It must be noted that Eq.~(\ref{eq:L_Anisotropy}) is only meaningful close to the interface. The application of this equation in the complete simulation domain can lead to spurious phenomena and can cause numerical instabilities. Du and Yu \cite{duVariationalConstructionAnisotropic2005} proposed a variational formulation to extrapolate the mobilities calculated at the interface to the whole domain. While this offers a systematic solution, it requires a significant number of numerical optimization steps adding to the computational complexity of the model. Alternatively, a cut-off order parameter value $\eta_\text{cut-off}$ can be defined which sets $L_\eta=\frac{L_{\eta}^0}{1+\beta}$ wherever $\eta \leq \eta_\text{cut-off}$, which is employed in this work.

\subsubsection{Auxiliary variables and numerical integration}\label{num-int}
The direct solution of Eq.~(\ref{eq:CH}) can promote numerical instabilities when evaluating the logarithmic terms in Eq.~(\ref{eq:reg-sol}) at low element compositions. From Fig.~\ref{fig:DF} it can be seen that very low concentration values can occur as dictated by the equilibrium condition. Therefore, we solve the Cahn-Hilliard equations using an auxiliary variable that is derived from the real compositions:

\begin{equation}
\label{eq:logarithmic-comps}
\begin{aligned}
Y_\text{Cu} &=  \ln \left( \frac{c_{\text{Cu}}}{1-c_{\text{Cu}}} \right),\\
Y_\text{Li} &=  \ln \left( \frac{c_\text{Li}}{1-c_\text{Li}} \right).
\end{aligned}
\end{equation}

In Eq.~(\ref{eq:logarithmic-comps}), $Y_\text{Cu}$ and $Y_\text{Li}$ denote the logarithmic mappings of the real compositions of Cu and Li, respectively. Substituting Eq.~(\ref{eq:logarithmic-comps}) into Eq.~(\ref{eq:CH}) leads to the following formalism:

\begin{equation}
\label{eq:CH-logarithmic-comps1}
\begin{aligned}
\frac{\partial Y_{\text{Cu}}}{\partial t} =&\nabla \cdot \left[M_{\text{Cu}} \nabla\left[V_m^{-1}\left[\mu_{\text{Cu}}-\mu_{\text{Al}}\right]\right]\right] -\frac{\partial\left[H(\boldsymbol{\eta})\left[v_{\text{Cu}}-c_{\text{Cu}}\right]\right]}{\partial t} -\left[ [1 - H(\boldsymbol{\eta})] \frac{\text{exp}({Y_{\text{Cu}}})}{\left[1+\text{exp}({Y_{\text{Cu}}})\right]^2} - 1 \right] \frac{\partial {Y_{\text{Cu}}}}{\partial t},\\
\frac{\partial Y_{\text{Li}}}{\partial t} =&\nabla \cdot \left[M_{\text{Li}} \nabla\left[V_m^{-1}\left[\mu_{\text{Li}}-\mu_{\text{Al}}\right]\right]\right] -\frac{\partial\left[H(\boldsymbol{\eta})\left[v_{\text{Li}}-c_{\text{Li}}\right]\right]}{\partial t} -\left[ [1 - H(\boldsymbol{\eta})] \frac{\text{exp}({Y_{\text{Li}}})}{\left[1+\text{exp}({Y_{\text{Li}}})\right]^2} - 1 \right] \frac{\partial {Y_{\text{Li}}}}{\partial t}.
\end{aligned}
\end{equation}

To enable the use of the semi-implicit spectral scheme, the terms $D^{*}_{Y_\text{Cu}} \nabla^2 Y_\text{Cu}$ and $D^{*}_{Y_\text{Li}} \nabla^2 Y_\text{Li}$, are introduced so the evolution equations can be written as:

\begin{equation}
\label{eq:CH-logarithmic-comps2}
\begin{aligned}
\frac{\partial Y_\text{Cu}}{\partial t} =& \nabla \cdot D^{*}_{Y_\text{Cu}}\nabla Y_\text{Cu} + f_{Y_\text{Cu}},\\
\frac{\partial Y_\text{Li}}{\partial t} =& \nabla \cdot D^{*}_{Y_\text{Li}}\nabla Y_\text{Li} + f_{Y_\text{Li}},
\end{aligned}
\end{equation}

and: 

\begin{equation}
\label{eq:CH-driving-force}
\begin{aligned}
f_{Y_\text{Cu}} &=\nabla \cdot \left[M_\text{Cu} \nabla\left[V_m^{-1}\left[\mu_\text{Cu}-\mu_\text{Al}\right]\right]\right] -\frac{\partial\left[H(\boldsymbol{\eta})\left[v_{\text{Cu}}-c_\text{Cu}\right]\right]}{\partial t} \\
&- \nabla \cdot D^{*}_{Y_\text{Cu}}\nabla Y_\text{Cu} - \left[ [1 - H(\boldsymbol{\eta})] \frac{\text{exp}({Y_\text{Cu}})}{\left[1+\text{exp}({Y_\text{Cu}})\right]^2} - 1 \right] \frac{\partial {Y_\text{Cu}}}{\partial t},\\
f_{Y_\text{Li}} &=\nabla \cdot \left[M_\text{Li} \nabla\left[V_m^{-1}\left[\mu_\text{Li}-\mu_\text{Al}\right]\right]\right] -\frac{\partial\left[H(\boldsymbol{\eta})\left[v_{\text{Li}}-c_\text{Li}\right]\right]}{\partial t} \\
&- \nabla \cdot D^{*}_{Y_\text{Li}}\nabla Y_\text{Li} - \left[ [1 - H(\boldsymbol{\eta})] \frac{\text{exp}({Y_\text{Li}})}{\left[1+\text{exp}({Y_\text{Li}})\right]^2} - 1 \right] \frac{\partial {Y_\text{Li}}}{\partial t},
\end{aligned}
\end{equation}

where $D^{*}_{Y}$ is a numerical value that is chosen to be $D^{*}_{Y_i}=10^3D_i$. We treat the diffusion terms in Eq.~(\ref{eq:CH-logarithmic-comps2}) implicitly while calculating the counterpart in Eq.~(\ref{eq:CH-driving-force}) explicitly together with the other terms. Thus, the evolution equations are discretized using the semi-implicit spectral method \cite{chenApplicationsSemiimplicitFourierspectral1998}. The Cahn-Hilliard evolution equations can be expressed in Fourier space via:

\begin{equation}
\label{CH-final}
\begin{aligned}
\hat{Y}_\text{Cu}^{n+1} &= \frac{\hat{Y}_\text{Cu}^n + \Delta t \hat{f}_{Y_\text{Cu}}^n}{1 + D^{*}_{Y_\text{Cu}} \Delta t \boldsymbol{\xi} \cdot \boldsymbol{\xi}},\\
\hat{Y}_\text{Li}^{n+1} &= \frac{\hat{Y}_\text{Li}^n + \Delta t \hat{f}_{Y_\text{Li}}^n}{1 + D^{*}_{Y_\text{Li}} \Delta t \boldsymbol{\xi} \cdot \boldsymbol{\xi}},
\end{aligned}
\end{equation}

and the discretized Allen-Cahn equation becomes:

\begin{equation}
\label{AC-final}
\begin{aligned}
\hat{\eta}_p^{n+1} = \frac{\hat{\eta}^n - \Delta t L_{\eta_p} \hat{f}_{\eta_p}^n}{1 + L_{\eta_p} \Delta t \boldsymbol{\kappa}_p \boldsymbol{\xi} \cdot \boldsymbol{\xi}},
\end{aligned}
\end{equation}

with 
\begin{equation}
\begin{aligned}
f_{\eta_p} = \frac{\partial h(\eta_p)}{\partial \eta_p} V_m^{-1} \Delta \mu^r\left(\mathbf{c}\right)+w \frac{\partial g^{d w}(\boldsymbol{\eta})}{\partial \eta_p}+\frac{\partial f_{e l}}{\partial \eta_p}.
\end{aligned}
\end{equation}

Equations (\ref{CH-final}) and (\ref{AC-final}) involve mobilities that vary both spatially and temporally. To accurately capture these variations, we employ protocols for varying mobilities as detailed in \cite{zhuCoarseningKineticsVariablemobility1999}. The model is implemented in Python using CuPy that allows for parallelization with GPUs, which significantly enhances computational efficiency and performance, as discussed by Boccardo et al. \cite{boccardoEfficiencyAccuracyGPUparallelized2023}.

\subsection{DFT}\label{DFT}
The calculation of elastic constants were performed using the Vienna Ab initio Simulation Package (VASP) version 6.3.0 \cite{kresseEfficiencyAbinitioTotal1996}. The projector augmented-wave (PAW) \cite{kresseUltrasoftPseudopotentialsProjector1999} method with the generalized gradient approximation (GGA-PBE) \cite{perdewGeneralizedGradientApproximation1996} was employed. The computational setup included an energy cutoff of 500~eV, and full relaxation of both the volume and shape of the crystal structures to the energy convergence of $10^{-6}$~eV and Hellmann-Feynman force tolerance of 0.01~eV/Angstrom. The elastic constants were calculated using the stress-strain method based on lattice distortion \cite{lepageSymmetrygeneralLeastsquaresExtraction2002} which is implemented in VASP. Monkhorst-Pack grids and gamma-centered grids were used for k-point mesh generation for Al and T$_1$, respectively. A k-point mesh of 25x25x25 for Al and 16x8x7 for T$_1$ was set for the calculations. Several experimentally proposed structures exist for T$_1$ phase~\cite{huangCrystalStructureStability1987,vansmaalenRefinementCrystalStructure1990,dwyerCombinedElectronBeam2011}. Kim et al.~\cite{kimFirstprinciplesStudyCrystal2018} thoroughly examined each proposed T$_1$ structures and suggested a novel structure for the T$_1$ phase in which the partial Li position was changed to z=0 and this atom no longer has partial occupancy. With a DFT energy that was lower than all of the experimentally suggested crystal structures, they discovered the best approximation of the disordered partial T$_1$ phase using the special quasi-random structure (SQS)~\cite{vandewalleEfficientStochasticGeneration2013} and cluster expansion (CE) approach~\cite{sanchezGeneralizedClusterDescription1984}, which is based on the Monte Carlo scheme. For the current investigation, modified T$_1$ structure was employed using the SQS scheme implemented in ICET~\cite{angqvistICETPythonLibrary2019}. The calculated elastic constants for T$_1$ and Al are given in Tabel~\ref{tbl:ParameterPF}.

\section{Results}\label{4}

In this section the results of the phase-field simulations are presented. It begins with investigations on one- and two-dimensional systems to explore the model predictions using the chemical potentials obtained from CALPHAD calculations and to investigate the effect of anisotropy in linear reaction rate to capture diffusion and interface controlled growth conditions. Subsequent sections present the phase-field simulations, focusing on the growth kinetics and morphological evolution of T\textsubscript{1} precipitates in a supersaturated Al-Cu-Li alloy. These simulations explore the effects of anisotropic interfacial energies and linear reaction rates on precipitate behavior, including multi-particle interactions.

\begin{table}[b]
\caption{Parameters for the phase-field model}
\centering
\begin{tabular}{p{0.2\linewidth}p{0.4\linewidth}p{0.3\linewidth}}
\toprule
Symbol & Description & Value \\
\midrule
$V_m$ & Molar volume  & 1.06 $\times$ 10\textsuperscript{-5} (m\textsuperscript{3}/mol) \\
$\lambda_b$ & Interface thickness of the broad interface & 1.0 (nm) \\
$w$ & Double-well height  & 1.32 $\times$ 10\textsuperscript{9} (J/m\textsuperscript{3}) \\
$L_{\eta_p}^0$ & Linear reaction rate for diffusion-controlled interface & $10^{-11}$ (m\textsuperscript{3}/Js) \\
$\beta$ & Anisotropy factor for the linear reaction rate& 10000 (-)\\
$\gamma_b, \gamma_p$ & Interfacial energy of broad and periphery interface & 0.110, 0.694 (J/m\textsuperscript{2}) \cite{AGUSTIANINGRUM2024174495}\\
$\kappa_b, \kappa_p$ & Gradient energy coefficient of broad and periphery interface & 1.65 $\times$ 10\textsuperscript{-10}, 1.04 $\times$ 10\textsuperscript{-9} (J/m) \\
${}^0D^{\alpha}_\text{Cu}, {}^0D^{\alpha}_\text{Li}$ & Diffusion pre-exponential in $\alpha$  & 6.5 $\times$ 10\textsuperscript{-5}, 3.5 $\times$ 10\textsuperscript{-5} (m\textsuperscript{2}/s) \cite{smithellsMetalsReferenceBook1992}\\
$Q^{\alpha}_\text{Cu}, Q^{\alpha}_\text{Li}$ & Diffusion activation energy in $\alpha$  & 136.0, 126.1 (kJ/mol) \cite{smithellsMetalsReferenceBook1992}\\
$\mathcal{C}^{\text{T}_1}_{11}, \mathcal{C}^{\text{T}_1}_{12}, \mathcal{C}^{\text{T}_1}_{13}, \mathcal{C}^{\text{T}_1}_{33}, \mathcal{C}^{\text{T}_1}_{44}$ & Coefficients of the elasticity tensor of T$_1$ & 165.2, 51.1, 30.0, 140.9, 62.5 (GPa) \\
$\mathcal{C}^{\alpha}_{11}, \mathcal{C}^{\alpha}_{12}, \mathcal{C}^{\alpha}_{44}$ & Coefficients of the elasticity tensor of $\alpha$ & 107.9, 62.9, 33.8 (GPa) \\
\bottomrule
\end{tabular}
\label{tbl:ParameterPF}
\end{table}

\subsection{1D simulations}
\begin{figure}[ht]
    \centering
    \includegraphics[width=0.4\textwidth]{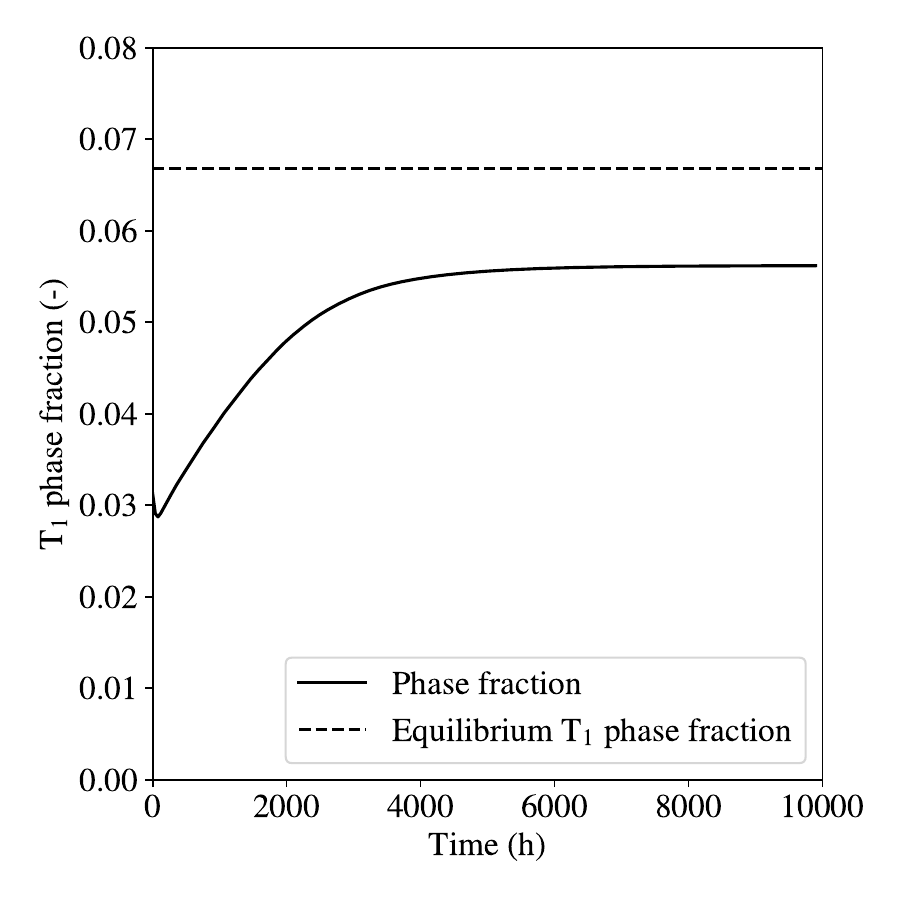}
    \caption{1D phase-field simulation of T$_1$ growth in a supersaturated Al-Cu-Li alloy.}
    \label{fig:OneD_Eq}
\end{figure}

In this section, the results of one-dimensional (1D) phase-field simulations are presented to establish an understanding of the growth kinetics of the T$_1$ precipitate in a supersaturated Al-Cu-Li alloy. The simulations were performed with a time step-size of $\Delta t = 1.42$~s, using 512 elements with a spatial step-size of $\Delta x = 0.5$~nm at an aging temperature of 155~$^{\circ}\mathrm{C}$. An initial seed with a length of $16\Delta x$ is placed. The domain has an initial composition of $c^{total}_{\text{Cu}}= 0.01, c^{total}_{\text{Li}}= 0.035$, while the solute composition, $c_{\text{Cu}}=0.009$ and $c_{\text{Li}}=0.034$, are initialized equally across the domain to maintain the total composition as defined in Eq.~(\ref{eq:total_comp}). The results, as shown in Fig.~\ref{fig:OneD_Eq}, indicate that according to the lever rule the predicted equilibrium volume fraction of T$_1$ precipitates under these conditions is 0.067. The volume fraction of T$_1$ precipitate gradually increases over time, converging to a value of 0.056. The discrepancy with respect to the lever rule prediction is attributed to interfacial energy contributions that lower the observed equilibrium phase fraction. Considering this effect, the obtained equilibrium phase fraction value lies within an expected range.

\begin{figure*}[ht]
    \centering
    \includegraphics[width=1\textwidth]{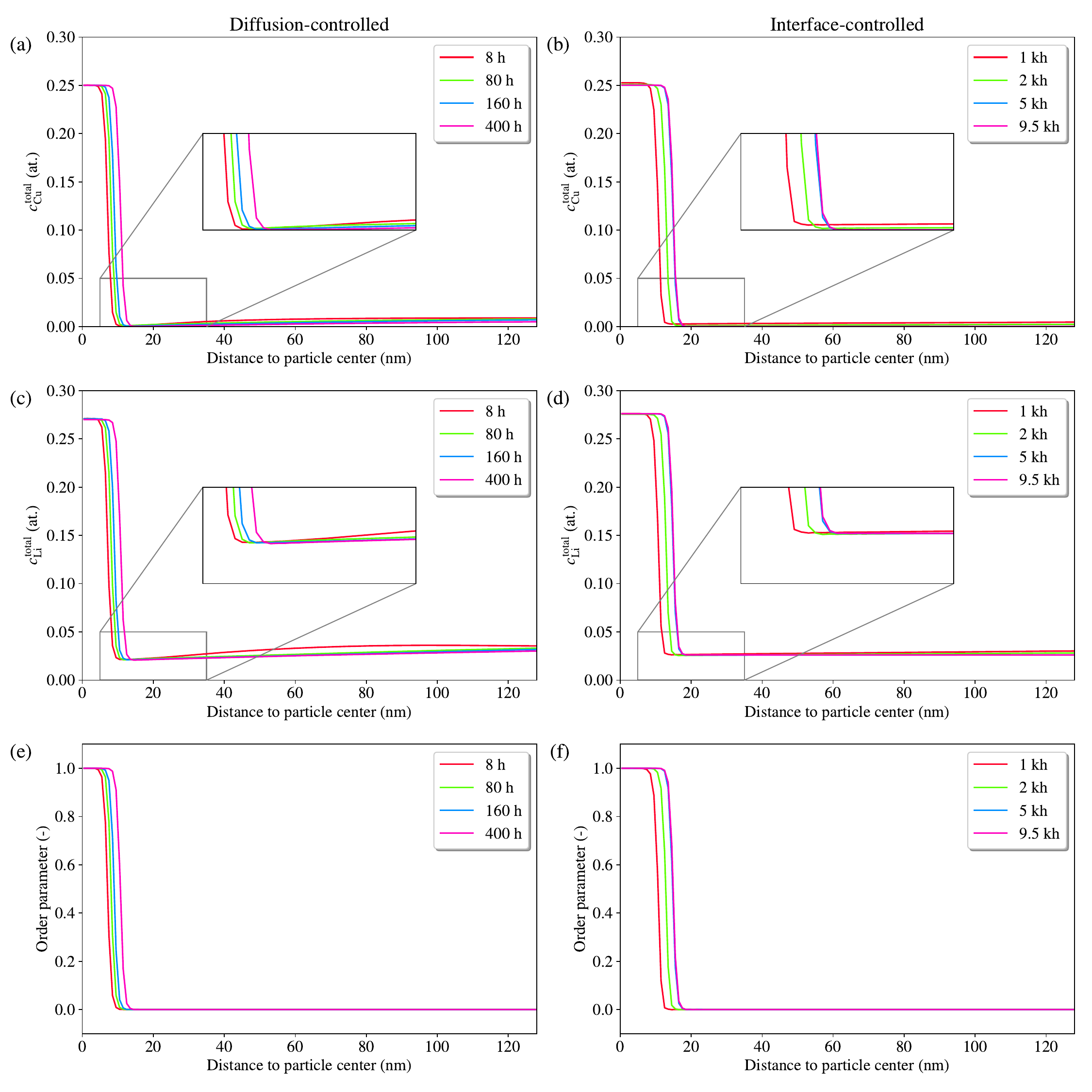}
    \caption{1D simulations of T$_1$ growth: (a) total Cu composition, (c) total Li composition, and (e) order parameter profile under diffusion-controlled conditions; (b) total Cu composition, (d) total Li composition, and (f) order parameter profile under interface-controlled conditions.}
    \label{fig:OneD_Diff_IntContr}
\end{figure*}

To further understand the growth dynamics, both diffusion-controlled and interface-controlled growth cases are analyzed. Since the linear reaction coefficient in multi-component systems is challenging to obtain directly, assumptions for the values of the basal and periphery coefficients are made. As discussed, experimental insights suggest that growth in the [111] direction is interface-controlled, while the lengthening in all the respective orthogonal directions occurs in a diffusion-controlled manner \cite{hauslerThickeningT1Precipitates2018}. Hence, values for both directions are assumed and ensured that they are in the suitable range knowing that possible validation with experimental results require a normalization of the time scale. The 1D simulations are conducted with assumed values of $L_{\eta_p}=10^{-11} \text{m}^3/\text{Js}$ for the diffusion controlled condition and $L_{\eta_p}=10^{-15}  \text{m}^3/\text{Js}$ for the interface controlled condition, setting the anisotropy coefficient $\beta$ to 10000. The simulation results for the two conditions are shown in Fig.~\ref{fig:OneD_Diff_IntContr}.

The diffusion-controlled growth profiles, Fig.~\ref{fig:OneD_Diff_IntContr} (a), (c), and (e), show a clear solute depletion zone for Cu and Li concentration. This depletion zone is evident from the sharp concentration gradient near the right side of precipitate-matrix interface, indicating limited solute availability in the vicinity of the growing precipitate. For Cu, the depletion zone widens as the aging time increases, with concentration profiles recorded at 8 hours, 80 hours, 160 hours, and 400 hours. Li, which diffuses significantly faster, as predicted by mobility values shown in Fig.~\ref{fig:Mobility}, also shows a visible solute depletion zone despite its rapid redistribution. Therefore, the particle growth rate in this case is limited and controlled by the amount of Cu and Li that diffuses to the particle interface.

The interface-controlled profiles are shown in Fig.~\ref{fig:OneD_Diff_IntContr}~(b),~(d), and (f). Under these conditions, the concentration profiles exhibit no pronounced solute depletion zones compared to diffusion-controlled growth, and the concentration distribution is more uniform across the precipitate-matrix interface. The equilibrium composition profile within the particle center is determined by the equilibrium conditions at 155~$^{\circ}\mathrm{C}$, see Fig.~\ref{fig:DF}. As the total Cu composition reaches the stoichiometric coefficient of 0.25, the Cu solute concentration becomes low, which requires an increase in total Li composition to maintain equilibrium. The result indicates that the chosen value of $L_{\eta_p}=10^{-15}  \text{m}^3/(\text{Js})$ is sufficient to guarantee interface-controlled growth. 

\begin{figure*}[ht]
    \centering
    \includegraphics[width=1\textwidth]{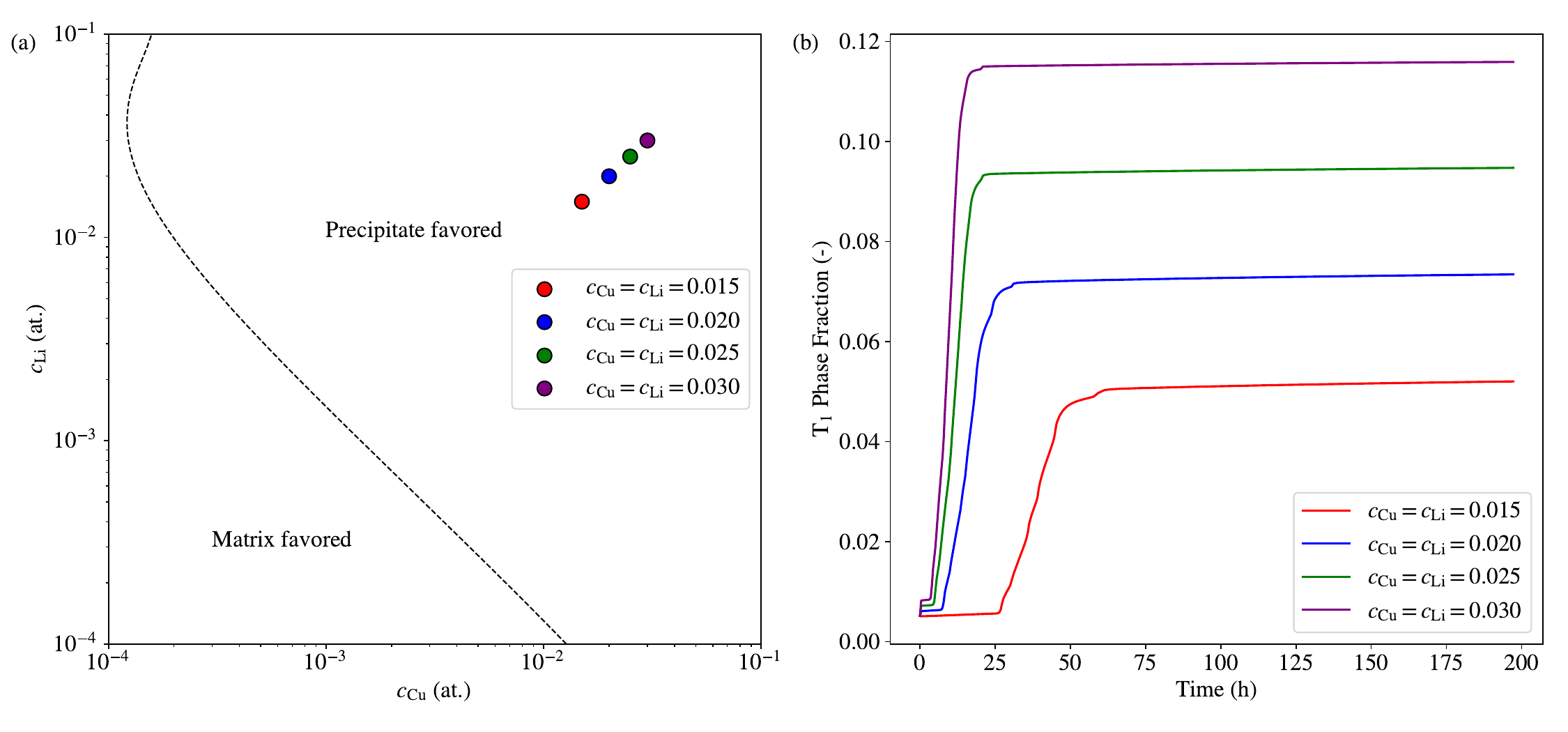}
    \caption{(a) Initial compositions of supersaturated matrix domains, and (b) the simulation results indicating growth depending on the amount of introduced solute.}
    \label{fig:OneD_OFFEq}
\end{figure*}

The compositional dynamics of precipitates are further illustrated by performing two-dimensional (2D) simulations in a domain with 512 elements in each dimension and a spatial step-size of $\Delta x = 0.5$~nm. A circular particle with a radius of $6\Delta x$ is placed in the center of the domain. Fig.~\ref{fig:OneD_OFFEq}~(a) presents the initial solute compositions of a supersaturated matrix. The matrix-favored and precipitate-favored regions are indicated, with equilibrium at 155~$^{\circ}\mathrm{C}$ shown as the transition boundary. As seen in Fig.~\ref{fig:OneD_OFFEq}~(b), the volume fraction of the precipitates in the supersaturated domain increases over time, indicating growth. Precipitates which are in supersaturated domains, with compositions above the equilibrium threshold, tend to grow as they seek to achieve equilibrium by absorbing solutes from the matrix to reach equilibrium. Naturally, the more available solute is introduced in the matrix the higher the equilibrium T\(_1\) phase fraction.

\subsection{Anisotropy simulations}

The strongly anisotropic shape of T$_1$ precipitates is attributed to contributions from elastic effects, anisotropy in interfacial energy and linear reaction rate. To quantify these effects, first a three-dimensional (3D) simulation on a $160\Delta x \times 160\Delta x \times 160\Delta x$ grid is performed with $\Delta x=1\mathrm{nm}$ to identify the effects of elasticity on precipitate growth. An initial spherical particle of $\eta_1=1$ is placed in the center with a diameter of $6\Delta x$. The simulation is conducted using an initial solute composition of $c_{\text{Cu}}=0.01$ and $c_{\text{Li}}=0.035$ at 155~$^{\circ}\mathrm{C}$, ensuring precipitate-favored thermodynamic conditions that promote particle growth. A reference case that neglects the elastic strain energy is used as a comparison to highlight the effects of individual contributions of the elastic anisotropy.  

\begin{figure*}[ht]
    \centering
    \includegraphics[width=0.5\textwidth]{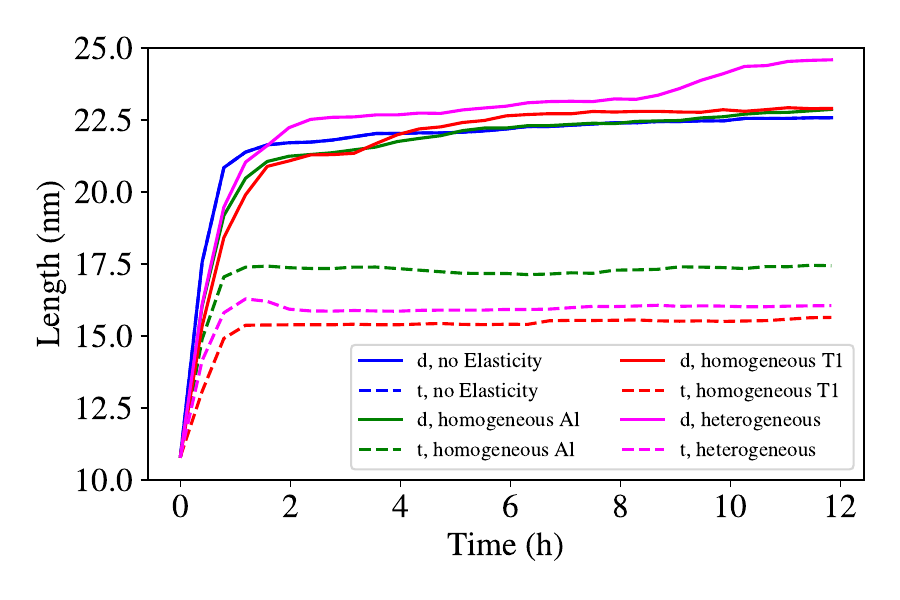}
    \caption{Evolution of the morphological parameters, diameter \(d\) and thickness \(t\), of the T\(_1\) precipitate over time, comparing scenarios with and without elastic effects. The result demonstrates that while the elastic effects slightly increase morphological anisotropy, they do not significantly alter the growth dynamics of the precipitate. Homogeneous Al or T\(_1\) refers to simulations, where the elasticity tensor consists only of pure Al or T\(_1\) contribution, respectively, for the whole domain. A heterogeneous condition implies the local interpolation of both elasticity tensors according to Eq.~(\ref{eq:Int_ETensor}).}
    \label{fig:elasticity}
\end{figure*}

The simulation results are presented in Fig.~\ref{fig:elasticity}, which show the evolution of the morphological parameters, diameter \(d\) and thickness \(t\), of the T\(_1\) precipitate over time considering elastic effects. From Fig.~\ref{fig:elasticity}, it can be observed that elastic effects do not significantly influence the growth dynamics of the precipitate. The morphological anisotropy increases when the elasticity tensor of T\(_1\) is considered in the full domain, leading to a precipitate diameter of 22.9 nm after 12 hours. In contrast, the thickness of the precipitate increases less over time, leading to a final value of 15.6 nm. It can be seen that in comparison to the reference case without elasticity, where the final diameter is 22.6 nm, the difference is not significantly pronounced.

\begin{figure*}[ht]
    \centering
    \includegraphics[width=0.7\textwidth]{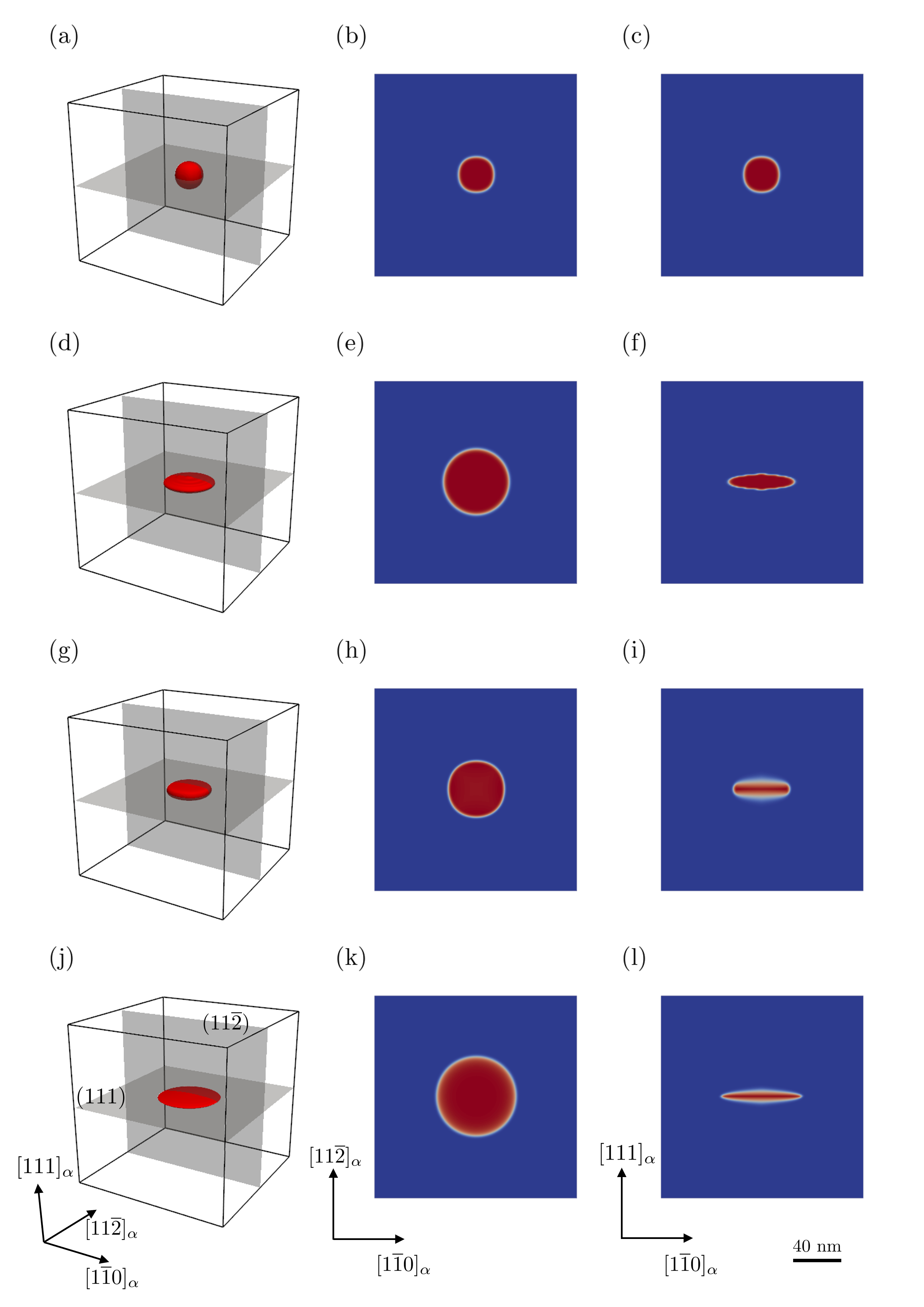}
    \caption{3D simulations of T$_1$ precipitate growth and order parameter plots of the $(111)$ and $(11\overline{2})$ planes for (a-c) isotropic interfacial energy and isotropic linear reaction rate, (d-f) anisotropic interfacial energy and isotropic linear reaction rate, (g-i) isotropic interfacial energy and anisotropic linear reaction rate, and (j-l) anisotropic interfacial energy and anisotropic linear reaction rate after 1.2 h.}
    \label{fig:Anisotropic}
\end{figure*}

Further 3D simulations were performed to investigate the effects of anisotropic interfacial energy and anisotropic linear reaction rates on the growth morphology of T$_1$ precipitates. Fig.~\ref{fig:Anisotropic} shows the results of the 3D simulations performed for a reference case with isotropic interfacial energy and cases with anisotropies in interfacial energy and linear reaction rate. Additionally, the plane sections with the highest anticipated respective anisotropy are shown, which correspond to the $(111)$ as well as the $(11\overline{2})$ planes.

\begin{figure*}[ht]
    \centering
    \includegraphics[width=1\textwidth]{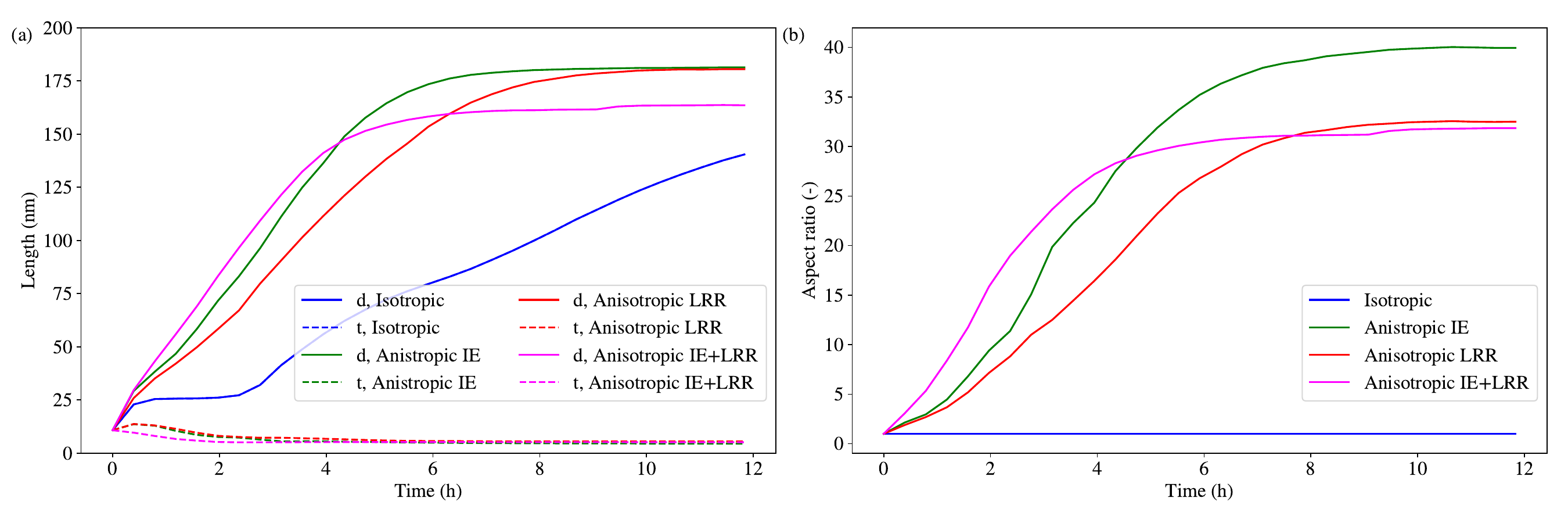}
    \caption{Quantitative analysis of T$_1$ precipitate growth: (a) Diameter, d, and thickness, t, measurements for isotropic interfacial energy (IE) and anisotropic linear reaction rate (LRR) conditions over time, and (b) aspect ratio evolution over time.}
    \label{fig:ShapeEvolution}
\end{figure*}

Under isotropic interfacial energy and isotropic linear reaction rate, Fig.~\ref{fig:Anisotropic} (a-c), the initial precipitate remains spherical as the isotropic conditions promote growth equally in all directions. For an anisotropic interfacial energy and isotropic linear reaction rate, Fig.~\ref{fig:Anisotropic} (d-f), the initial precipitate transforms into a flat shape with an ellipsoidal morphology. From Fig.~\ref{fig:ShapeEvolution} it can bee seen that the aspect ratio of a growing particle under anisotropic interfacial energy converges to 39.9. When considering isotropic interfacial energy and anisotropic linear reaction rate, Fig.~\ref{fig:Anisotropic} (g-i), the precipitate evolution results in a shape that exhibits a lower thickness during early aging stages comparing the effects of anisotropic interfacial energy. This is evident in the $(111)$ and $(11\overline{2})$ plane sections, Fig.~\ref{fig:Anisotropic} (j-l), where the precipitate is significantly elongated, indicating the substantial influence of anisotropic reaction kinetics on precipitate morphology. When combining anisotropy in interfacial energy and linear reaction rate, the particle grows to an aspect ratio of 31.87.

\subsection{Multi-particle simulations}
\begin{figure*}[ht]
    \centering
    \includegraphics[width=0.8\textwidth]{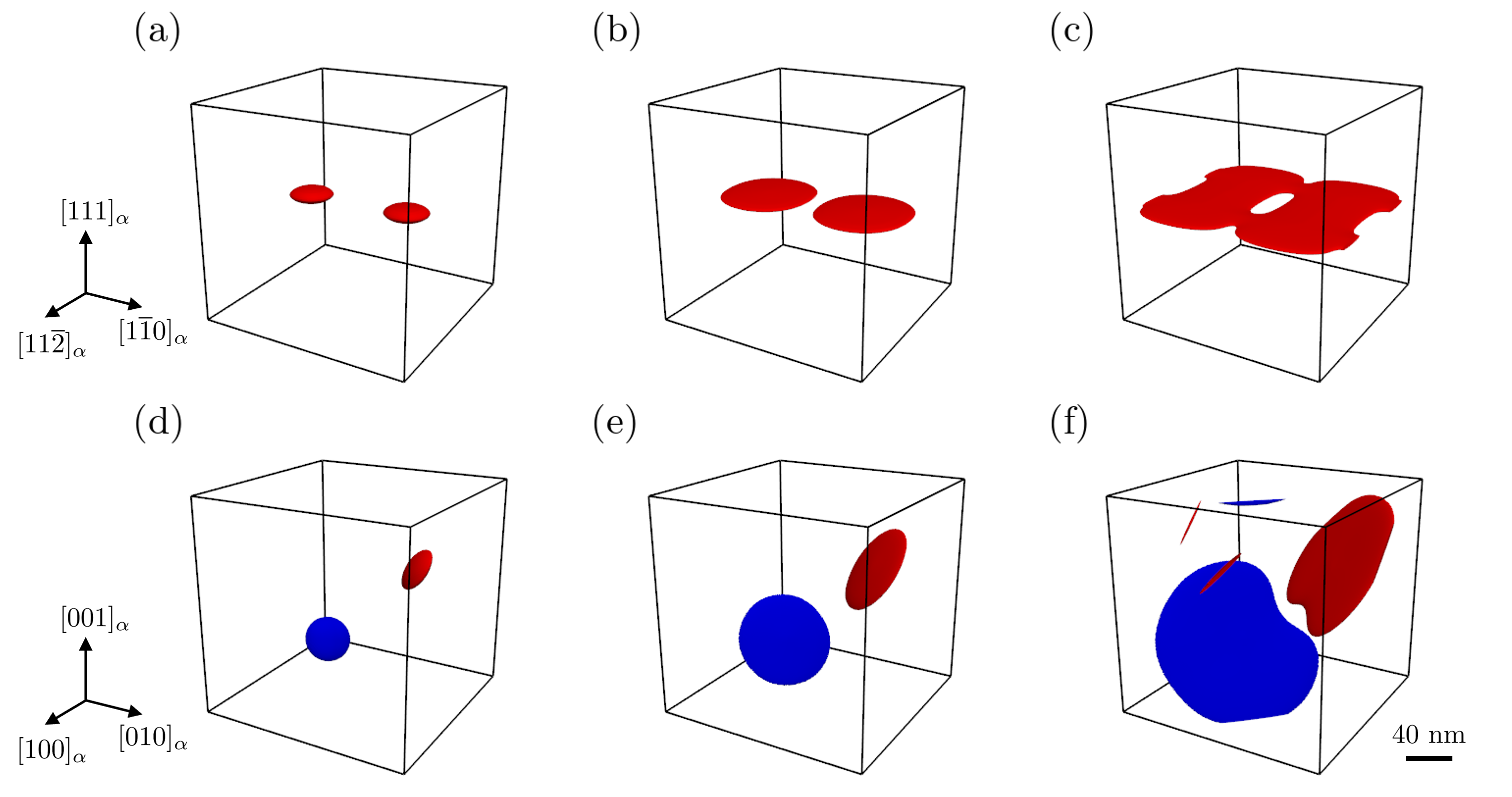}
    \caption{Interaction of two T$_1$ precipitates under conditions of anisotropic interfacial energy and anisotropic linear reaction rates. (a-c) represent a time series of 0.4, 1.6 and 3.2 h for particles with the same order parameter, and (d-f) for particles with different order parameters.}
    \label{fig:TwoParticle}
\end{figure*}

To investigate the interaction of particles influenced by anisotropic interfacial energy and anisotropic linear reaction rates, simulations are conducted involving two precipitates, each initially configured with a spherical morphology and a diameter of $6\Delta x$. The progression over time (0.4, 1.6, and 3.2 hours) is depicted in Fig.~\ref{fig:TwoParticle}. Fig.~\ref{fig:TwoParticle}~(a-c) demonstrates the temporal evolution of two particles with the same order parameter, whereas Fig.~\ref{fig:TwoParticle}~(d-f) illustrates the interaction between two particles with differing order parameter, representing two different particle variants. 

In Fig.~\ref{fig:TwoParticle}~(a-c), the particles initially exhibit an ellipsoidal shape. Due to the influence of anisotropic interfacial energy and anisotropic linear reaction rates, they quickly begin to elongate. The anisotropy promotes directional growth, causing the particles to grow towards each other and eventually merge. This merging behavior can be explained by the double-well potential formalism, where particles with the same order parameter are energetically favorable to coalesce, thereby reducing the system's overall free energy.

Conversely, particles with different order parameters, see Fig.~\ref{fig:TwoParticle}~(d-f), experience a repulsive interaction. In this scenario, the double-well potential prevents the merging of particles with differing order parameters, as it is energetically unfavorable for them to coexist. This repulsion inhibits their growth and leads to reduced particle growth rates. The particles retract from each other, maintaining distinct identities rather than merging, which significantly influences their morphological evolution. The differing behaviors observed in the two cases can be attributed to the interaction energies. For particles with the same order parameter, the interfacial energy and reaction rates drive the particles to merge, minimizing the system's free energy. 

\begin{figure*}[ht]
    \centering
    \includegraphics[width=1\textwidth]{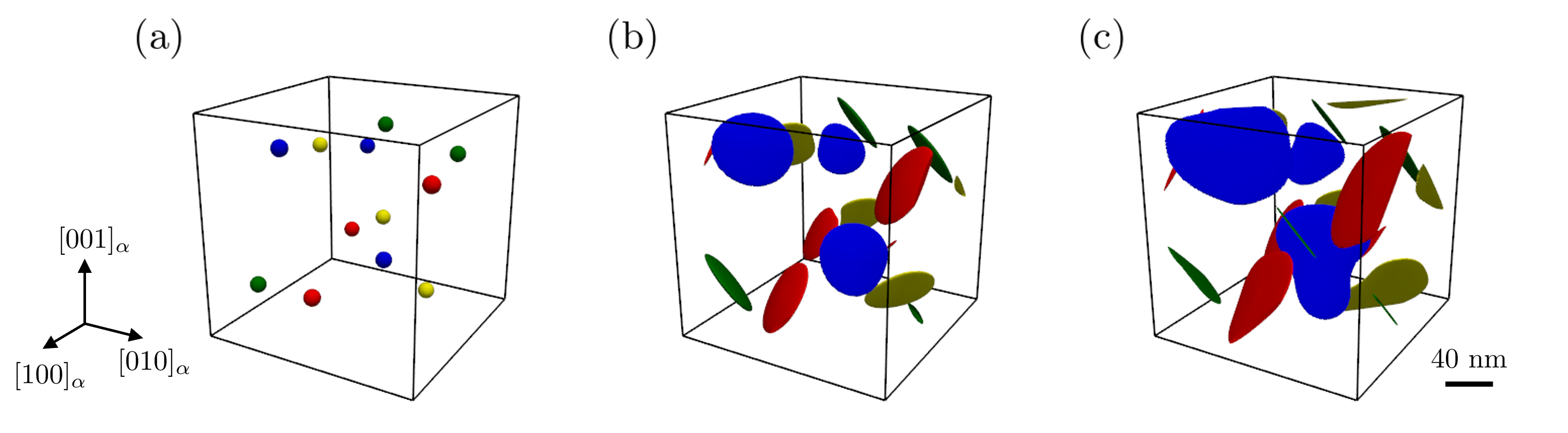}
    \caption{Time evolution of a multi-particle system with twelve precipitates, showcasing the spatial distribution and morphological of (a) the initial setup changes over time (b) after 1.2 h and (c) after 3.15 h.}
    \label{fig:Multiparticle}
\end{figure*}

The simulations are extended to model a multi-particle system containing twelve particles, with three particles of each variant, randomly distributed. The initial particles have a diameter of $6\Delta x$, which can be seen in Fig.~\ref{fig:Multiparticle}~(a). This setup corresponds to a precipitate number density of $2.9 \times 10^{21} \text{m}^{-3}$ which matches well with experimentally observed values in similar alloys \cite{hauslerThickeningT1Precipitates2018, dorinQuantificationModellingMicrostructure2014}. The results of these simulations over time are illustrated in Fig.~\ref{fig:Multiparticle}~(b,c).

Initially, the particles are uniformly distributed and exhibit a plate-like morphology. As the simulation progresses, the particles interact due to the effects of anisotropic interfacial energy and anisotropic linear reaction rates. These interactions lead to significant morphological changes, including elongation and coalescence of particles with the same order parameter, and repulsion among particles with different order parameters. A pronounced lengthening process can be observed, where the particles reach a diameter of up to 80 nm while all the particles are at least 30 nm long. Particles with different order parameters interact with each other, as observed for two precipitates in Fig.~\ref{fig:TwoParticle}~(d-f), and exhibit limited growth and maintain higher aspect ratios due to the repulsive interactions that inhibit their coalescence. The aspect ratio shows a trend of thickening for some particles, indicating that while the particles continuously grow in size, their interaction can promote thickening for selective particles. This demonstrates the complex dynamics of multi-particle interactions under anisotropic conditions. The anisotropic interfacial energy and linear reaction rate not only influence the growth and morphology of individual particles but also dictate the collective behavior of the particles.

\section{Conclusion}\label{5}

A multi-component phase-field model was presented that successfully simulates the precipitation behavior of T$_1$ precipitates. The simulations highlight the significant influence of anisotropic interfacial energy and anisotropic linear reaction rates on the morphological evolution of precipitates. The main conclusions can be summarized as follows:

\begin{itemize}
\item 1D simulations highlight the differences between diffusion-controlled and interface-controlled growth. It is observed that the availability of solute Cu and Li in the vicinity of the periphery $\alpha/\text{T}_1$ interface acts as the limiting factor for the lengthening process of precipitates.
\item The role of elasticity in morphological anisotropy is found to be negligible under the examined conditions.
\item The combined effects of interfacial energy anisotropy and linear reaction rates are crucial in ensuring nearly constant thickness and diffusion-controlled lengthening, leading to realistic particle morphologies.
\item Multi-particle simulations further demonstrate the dynamics of precipitate interactions. Particles sharing the same order parameter tend to coalesce, thereby promoting growth and reducing the overall system energy. Conversely, particles with differing order parameters exhibit repulsive interactions, inhibiting coalescence and resulting in distinct particle morphologies.
\end{itemize}

The findings underscore the importance of considering anisotropic effects in the phase-field modeling of precipitate evolution in Al-Cu-Li alloys. This enhanced model provides insights into the mechanisms of precipitate coalescence and thickening, which are essential for the design and optimization of age-hardenable Al alloys.

\printcredits

\section*{Declaration of competing interest}
The authors declare that they have no known competing financial interests or personal relationships that could have appeared to influence the work reported in this paper.

\section*{Data availability}
The obtained data from this research will be made available on ZENODO.

\section*{Code availability}
The codes from this research is available on github (\url{https://github.com/alisafi96/StoichiometricPF_T1}).

\section*{Acknowledgements}
This project has received funding from the European Research Council (ERC) under the European Union’s Horizon 2020 research and innovation programme (grant  agreement No 101001567).

\bibliographystyle{elsarticle-num}

\bibliography{MyBib}

\appendix

\numberwithin{equation}{section}
\numberwithin{figure}{section}
\numberwithin{table}{section}

\section{Derivation of bulk free energy density}\label{app1}

In the following the bulk free energy density, as shown in Eq.~(\ref{eq:bulk}), is derived starting from the proposed reaction in Eq.~(\ref{eq:reaction}). For reasons of simplicity, we will equivalently denote $\text{Al}_{v_\text{Al}}\text{Cu}_{v_\text{Cu}}\text{Li}_{v_\text{Li}}$ as T$_1$. The total Gibbs free energy, G, of the system can be described as:
\begin{equation}
\begin{aligned}
 G =& \mu_\text{Al}N_\text{Al} + \mu_\text{Cu}N_\text{Cu} +  \mu_\text{Li}N_\text{Li}+ \mu^{\text{T}_{\text{1}}}N_{\text{T}_1},
\end{aligned}
\end{equation}

where $\mu_i$ denotes the chemical potential of element $i$ and $N_i$ represents the amount of substance. The total amount of substance, $N_{i}^{\text{tot}}$, can be described via:
\begin{equation}
\begin{aligned}
N_{i}^{\text{tot}} &= N_{i} + v_{i}N_{\text{T}_1}, 
\end{aligned}
\end{equation}

Naturally, the infinitesimal change in Gibbs free energy is defined by the following relation:
\begin{equation}
\begin{aligned}
\text{d}G =& \mu_\text{Al}\,\text{d}N_\text{Al} + \mu_\text{Cu}\,\text{d}N_\text{Cu} +  \mu_\text{Li}\,\text{d}N_\text{Li} + \mu^{\text{T}_{\text{1}}}\,\text{d}N_{\text{T}_1},
\end{aligned}
\end{equation}

and the free energy density, $f_{\text{bulk}}$, reads:
\begin{equation}
\label{eq:fbulk}
\begin{aligned}
f_{\text{bulk}} = \frac{\partial G}{\partial V} =& \mu_\text{Al}m_\text{Al} + \mu_\text{Cu}m_\text{Cu} +  \mu_\text{Li}m_\text{Li}+ \mu^{\text{T}_{\text{1}}}m_{\text{T}_1},
\end{aligned}
\end{equation}

where $m_{i}$ denotes the molar density of element $i$. The molar densities can be related to $c_i$ and $\eta$ in the following manner:
\begin{equation}
\label{eq:fractions}
\begin{aligned}
\frac{m^s}{m} = [1-\eta] ; \; \frac{m_i}{m^s} = c_i; \; \frac{m_\mathrm{\text{T}_1}}{m} = \eta,
\end{aligned}
\end{equation}

where $m^s$ represent the amount of solid solution and $m$ is the total number of atoms per volume, i. e. the reciprocal of the molar volume $V_m$. Since the proposed reaction in Eq.~(\ref{eq:reaction}) is heterogeneous, $c_i$ is defined as $\frac{m_i}{m^s}$ instead of $\frac{m_i}{m}$. Eq.~(\ref{eq:fbulk}) can now also be written as:
\begin{equation}
\label{eq:fbulk2}
\begin{aligned}
f_{\text{bulk}} =& [1-H(\boldsymbol{\eta})] \frac{m}{m^s} [\mu_\text{Al}m_\text{Al} + \mu_\text{Cu}m_\text{Cu} +  \mu_\text{Li}m_\text{Li}]+ \mu^{\text{T}_{\text{1}}}H(\boldsymbol{\eta})m,
\end{aligned}
\end{equation}

where $\eta$ is substituted by $H(\boldsymbol{\eta})$ to allow for a smooth interpolation between all variants described in section \ref{2} of the manuscript. This simplifies Eq.~(\ref{eq:fbulk3}) to:

\begin{equation}
\label{eq:fbulk3}
\begin{aligned}
f_{\text{bulk}} =& \frac{1}{V_m} \left[ [1-H(\boldsymbol{\eta})][\mu_\text{Al}c_\text{Al} + \mu_\text{Cu}c_\text{Cu} +  \mu_\text{Li}c_\text{Li}]+ \mu^{\text{T}_{\text{1}}}H(\boldsymbol{\eta}) \right], \\
=& \frac{1}{V_m} \left[  \left[1 - H(\boldsymbol{\eta}) \right] \mu^{\alpha} (\mathbf{c}) + H(\boldsymbol{\eta}) \mu^{\text{T}_{\text{1}}} \right].
\end{aligned}
\end{equation}

Eq.~(\ref{eq:fbulk3}) allows the smooth interpolation of the bulk free energy density between composition dependent contributions of the $\alpha$-Al phase and the chemical potential of the stoichiometric T$_1$ compound.

\section{Derivation of evolution equations}\label{app2}

The Cahn-Hilliard and Allen-Cahn evolution equations for the stoichiometric reaction shown in Eq.~(\ref{eq:reaction}) are derived starting from the definition of the bulk free energy density as given in Eq.~(\ref{eq:fbulk}). The molar density, $m_{i}$, is related to the total molar density and the total molar volume as:
\begin{equation}
\begin{aligned}
m=\frac{1}{V^m}=&m_{\mathrm{Al}}+m_{\mathrm{Cu}} + m_{\mathrm{Li}}+m_\mathrm{\text{T}_1},\\
=&m_{\mathrm{Al}}^{\text{tot}}+m_{\mathrm{Cu}}^{\text{tot}} +m_{\mathrm{Li}}^{\text{tot}}.
\end{aligned}
\end{equation}

From mass conservation the following relation can be established:
\begin{equation}
\label{eq:masscons}
\begin{aligned}
\text{d}m_{\mathrm{Al}}^{\text{tot}}+\text{d}m_{\mathrm{Cu}}^{\text{tot}} +\text{d}m_{\mathrm{Li}}^{\text{tot}} = 0.
\end{aligned}
\end{equation}

The total molar density, $m_{i}^{\text{tot}}$, of element $i$ can be described by the amount in solid solution and stoichiometric contribution from the compound phase:
\begin{equation}
\label{eq:tot}
\begin{aligned}
m_{i}^{\text{tot}} = m_{i} + v_{i}\,\text{d}m_\mathrm{\text{T}_1}.
\end{aligned}
\end{equation}

By combining the relation established in Eq.~(\ref{eq:tot}), the infinitesimal change of Eq.~(\ref{eq:fbulk}) can be expressed by:
\begin{equation}
\begin{aligned}
\text{d}f_{\text{bulk}} =& \mu_\text{Al}[\text{d}m^{\text{tot}}_\text{Al} - v_\text{Al}\,\text{d}m_\mathrm{\text{T}_1}]
+ \mu_\text{Cu}[\text{d}m^{\text{tot}}_\text{Cu} - v_\text{Cu}\,\text{d}m_\mathrm{\text{T}_1}] 
+ \mu_\text{Li}[\text{d}m^{\text{tot}}_\text{Li} - v_\text{Li}\,\text{d}m_\mathrm{\text{T}_1}] + \mu^{\text{T}_{\text{1}}}\,\text{d}m_\mathrm{\text{T}_1},\\
=& \mu_\text{Al}\,\text{d}m^{\text{tot}}_\text{Al} + \mu_\text{Cu}\,\text{d}m^{\text{tot}}_\text{Cu} + \mu_\text{Li}\,\text{d}m^{\text{tot}}_\text{Li} 
+ [\mu^{\text{T}_{\text{1}}} - v_\text{Al}\mu_\text{Al} - v_\text{Cu}\mu_\text{Cu} - v_\text{Li}\mu_\text{Li}]\,\text{d}m_\mathrm{\text{T}_1},
\end{aligned}
\end{equation}

with $(\mu_\text{Cu} - \mu_\text{Al})$ and $(\mu_\text{Li} - \mu_\text{Al})$ being the diffusion potentials of Cu and Li. By employing the mass conservation described in Eq.~(\ref{eq:masscons}), we can further simplify $\text{d}f_{\text{bulk}}$ to:
\begin{equation}
\begin{aligned}
\text{d}f_{\text{bulk}} = & [\mu_\text{Cu} - \mu_\text{Al}]\,\text{d}m_\text{Cu}^{\text{tot}} + [\mu_\text{Li} - \mu_\text{Al}]\,\text{d}m_\text{Li}^{\text{tot}} +[\mu^{\text{T}_{\text{1}}} - v_\text{Al}\mu_\text{Al} - v_\text{Cu}\mu_\text{Cu} - v_\text{Li}\mu_\text{Li}]\,\text{d}m_\mathrm{\text{T}_1},
\end{aligned}
\end{equation}

which can also be expressed as:
\begin{equation}
\begin{aligned}
\text{d}f_{\text{bulk}} =& m[\mu_\text{Cu} - \mu_\text{Al}]\,\text{d}c_\text{Cu}^{\text{tot}} + m[\mu_\text{Li} - \mu_\text{Al}]\,\text{d}c_\text{Li}^{\text{tot}}+m[\mu^{\text{T}_{\text{1}}} - v_\text{Al}\mu_\text{Al} - v_\text{Cu}\mu_\text{Cu} - v_\text{Li}\mu_\text{Li}]\,\text{d}\eta,
\end{aligned}
\end{equation}

with $c_{i}^{\text{tot}}$ being the total concentration of element $i$ and $\eta$ represents the extent of the reaction described in Eq.~(\ref{eq:reaction}). The diffusion driving force can be expressed as:
\begin{equation}
\begin{aligned}
\frac{\partial f_{\text{bulk}}}{\partial c_\text{Cu}^{\text{tot}}} =& m[\mu_\text{Cu} - \mu_\text{Al}], \\
\frac{\partial f_{\text{bulk}}}{\partial c_\text{Li}^{\text{tot}}} =& m[\mu_\text{Li} - \mu_\text{Al}],
\end{aligned}
\end{equation}

and the derivative of the bulk free energy density with respect to the order parameter is defined as:
\begin{equation}
\begin{aligned}
\frac{\partial f_{\text{bulk}}}{\partial \eta} =& m[\mu^{\text{T}_{\text{1}}} - v_\text{Al}\mu_\text{Al} - v_\text{Cu}\mu_\text{Cu} - v_\text{Li}\mu_\text{Li}],\\
=& m \Delta \mu^r,
\end{aligned}
\end{equation}

with $\Delta \mu^r$ being the reaction driving force. The Allen-Cahn evolution equation and the the Cahn-Hilliard equation can now be constructed as follows:
\begin{equation}
\label{eq:AC_app}
\begin{aligned}
\frac{\partial \eta}{\partial t} =& -L_{\eta} \left[\frac{\partial f_{\text{bulk}}}{\partial \eta}\right] =-L_{\eta} m\Delta \mu^r,
\end{aligned}
\end{equation}

and:
\begin{equation}
\label{eq:CH_app}
\begin{aligned}
\frac{\partial c_{i}^{\text{tot}}}{\partial t} =& \nabla \cdot \left[M_{i}\nabla \frac{\partial f_{\text{bulk}}}{\partial c_{i}^{\text{tot}}} \right],
\end{aligned}
\end{equation}

with $M_i$ being the atomic/chemical mobility of element $i$. It must be noted that the chemical potentials are usually expressed in terms of $c_i$ instead of $c_i^{\text{tot}}$. Therefore, we seek to adjust Eq.~(\ref{eq:AC_app}) and Eq.~(\ref{eq:CH_app}) to ensure consistency with CALPHAD-derived thermodynamical data. $c_i^{\text{tot}}$ can be further defined as:
\begin{equation}
\label{eq:citot}
\begin{aligned}
c_{i}^{\text{tot}} =&  \frac{m_{i}^{\text{tot}}}{m} = \frac{m_{i} + v_i m_\mathrm{\text{T}_1}}{m} = \frac{m^s}{m} \frac{m_i}{m^s} + v_i \frac{m_\mathrm{\text{T}_1}}{m}.
\end{aligned}
\end{equation}

Using the relations established in Eq.~(\ref{eq:fractions}), Eq.~(\ref{eq:citot}) simplifies to:
\begin{equation}
\begin{aligned}
c_{i}^{\text{tot}} =& [1-\eta]c_{i} + v_{i}\eta.
\end{aligned}
\end{equation}

By applying the product rule, the time derivative can be expressed as:
\begin{equation}
\label{eq:citot-simp}
\begin{aligned}
\frac{\partial c_{i}^{\text{tot}}}{\partial t} =& \frac{\partial [1-\eta] c_{i}}{\partial t} + \frac{\partial v_{i}\eta}{\partial t},\\
=& \frac{\partial c_{i}}{\partial t} \underbrace{- \eta \frac{\partial c_{i}}{\partial t} + [v_{i} - c_{i}] \frac{\partial \eta}{\partial t}}_{\frac{\partial \eta[v_{i} - c_{i}]}{\partial t}},\\
=&\frac{\partial c_{i}}{\partial t} + \frac{\partial \eta[v_{i} - c_{i}]}{\partial t} .
\end{aligned}
\end{equation}

Combining Eq.~(\ref{eq:CH_app}) with Eq.~(\ref{eq:citot-simp}) allows a new definition of the Cahn-Hilliard equations that solve for $c_{i}$ instead of $c_{i}^{\text{tot}}$. For a system with multiple order parameters it is useful to express the source terms using the interpolation function $H(\boldsymbol{\eta})$:
\begin{equation}
\begin{aligned}
\frac{\partial c_\text{Cu}}{\partial t} =& \nabla \cdot \left[ V_m^{-1} M_\text{Cu}\nabla \left[ \mu_\text{Cu} - \mu_\text{Al} \right]\right] - \left[ \frac{\partial H(\boldsymbol{\eta})[v_\text{Cu} - c_\text{Cu}]}{\partial t}\right],
\end{aligned}
\end{equation}

\begin{equation}
\begin{aligned}
\frac{\partial c_\text{Li}}{\partial t} =& \nabla \cdot \left[ V_m^{-1}M_\text{Li}\nabla \left[ \mu_\text{Li} - \mu_\text{Al} \right]\right] - \left[ \frac{\partial H(\boldsymbol{\eta})[v_\text{Li} - c_\text{Li}]}{\partial t}\right].
\end{aligned}
\end{equation}

In summary, the source terms allow to describe the evolution equation in terms of the element composition in solid solution as required by most CALPHAD databases. This enables a consistent description with the Allen-Cahn equation and enforces mass conservation.

\section{Data}\label{app4}

\begin{table}[pos=h!]
\centering
\caption{Coefficients for Al, Cu, and Li for the chemical potential polynomial of the pure elements.}\label{tbl:Gibbs_alpha_coeff}
\setlength\tabcolsep{6pt}
\begin{tabular}{cccccccc}
\toprule
Element $i$ & $A^{0}_i$ & $B^{0}_i$ & $C^{0}_i$ & $D^{0}_i$ & $E^{0}_i$ & $F^{0}_i$ & reference\\
\midrule
$i=1$ (\text{Al}) & $-7976.15$ & $137.093$ & $-0.00188466$ & $-0.000000877664$ & $-24.3672$ & $74092$ & \cite{saunders1998cost}\\
$i=2$ (\text{Cu}) & $-7770.458$ & $130.485$ & $-0.00265684$ & $0.000000129223$ & $-24.1124$ & $52478$ & \cite{saunders1998cost} \\
$i=3$ (\text{Li}) & $-10583.817$ & $217.637$ & $0.0354669$ & $-0.0000198698$ & $-38.9405$ & $159994$ & \cite{saunders1998cost}\\
\bottomrule
\end{tabular}
\end{table}

\begin{table}[pos=h!]
\centering
    \caption{Values of binary interaction coefficient ${}^{k}L_{i j}^{\alpha}$.}\label{tbl:binary_interaction}
    \setlength\tabcolsep{6pt}
    \begin{tabular}{ccccc}
    \toprule
    Elements & $k=0$ & $k=1$ & $k=2$ & reference\\
    \midrule
    $i=1 (\text{Al}), j =2 (\text{Cu})$ & $-53520 + 2T$ & $38590 - 2T$ & $1170$ & \cite{kroupaThermodynamicReassessmentBinary2021}\\
    $i=1 (\text{Al}), j =3 (\text{Li})$ & $-27000.0 + 8.0T$ & $1 \times 10^{-6}$ & $3000.0 + 0.1T$ & \cite{azzaThermodynamicDescriptionAluminumLithium2015}\\
    $i=2 (\text{Cu}), j =3 (\text{Li})$ & $2750 + 13.0T$ & $-1000$ & $0.0$  & \cite{liThermodynamicAssessmentCu2016}\\
    \bottomrule
    \end{tabular}
\end{table}

\begin{table}[pos=h!]
\centering
    \caption{Rotation matrix for each T\textsubscript{1}-variant}\label{Rotation_matrix}
    \begin{tabular}{ccc}
    \toprule
    Variant & Transformed principal axes & Rotation matrix \\
    \midrule
    1 & $\begin{matrix}
    [\overline{1} 1 0]\rightarrow [1 0 0]\\
    [1 1 1]\rightarrow [0 1 0] \\
    [1 1 \overline{2}]\rightarrow [0 0 1] \\
    \end{matrix}$ & $\begin{pmatrix}
    \frac{-1}{\sqrt{2}} & \frac{1}{\sqrt{3}} & \frac{1}{\sqrt{6}} \\
    \frac{1}{\sqrt{2}} & \frac{1}{\sqrt{3}} & \frac{1}{\sqrt{6}} \\
    0 & \frac{1}{\sqrt{3}} & \frac{-2}{\sqrt{6}}
    \end{pmatrix}$ \\
    2 & $\begin{matrix}
    [\overline{1} \overline{1} 0]\rightarrow [1 0 0]\\
    [\overline{1} 1 1]\rightarrow [0 1 0] \\
    [\overline{1} 1 \overline{2}]\rightarrow [0 0 1] \\
    \end{matrix}$ & $\begin{pmatrix}
    \frac{-1}{\sqrt{2}} & \frac{-1}{\sqrt{3}} & \frac{-1}{\sqrt{6}} \\
    \frac{-1}{\sqrt{2}} & \frac{1}{\sqrt{3}} & \frac{1}{\sqrt{6}} \\
    0 & \frac{1}{\sqrt{3}} & \frac{-2}{\sqrt{6}}
    \end{pmatrix}$ \\
    3 & $\begin{matrix}
    [1 1 0]\rightarrow [1 0 0]\\
    [1 \overline{1} 1]\rightarrow [0 1 0] \\
    [1 \overline{1} \overline{2}]\rightarrow [0 0 1] \\
    \end{matrix}$ & $\begin{pmatrix}
    \frac{1}{\sqrt{2}} & \frac{1}{\sqrt{3}} & \frac{1}{\sqrt{6}} \\
    \frac{1}{\sqrt{2}} & \frac{-1}{\sqrt{3}} & \frac{-1}{\sqrt{6}} \\
    0 & \frac{1}{\sqrt{3}} & \frac{-2}{\sqrt{6}}
    \end{pmatrix}$ \\
    4 & $\begin{matrix}
    [1 \overline{1} 0]\rightarrow [1 0 0]\\
    [1 1 \overline{1}]\rightarrow [0 1 0] \\
    [1 1 2]\rightarrow [0 0 1] \\
    \end{matrix}$ & $\begin{pmatrix}
    \frac{1}{\sqrt{2}} & \frac{1}{\sqrt{3}} & \frac{1}{\sqrt{6}} \\
    \frac{-1}{\sqrt{2}} & \frac{1}{\sqrt{3}} & \frac{1}{\sqrt{6}} \\
    0 & \frac{-1}{\sqrt{3}} & \frac{2}{\sqrt{6}}
    \end{pmatrix}$ \\
    \bottomrule
    \end{tabular}
\end{table}

\end{document}